\documentclass[sigconf]{acmart}

\usepackage{wrapfig}
\usepackage{multirow}
\usepackage{subfig}
\usepackage[utf8]{inputenc}
\usepackage[T1]{fontenc}
\usepackage[figuresright]{rotating}
\usepackage{makecell}
\usepackage{multirow}
\usepackage{url}
\usepackage{float}
\usepackage{rotating} 
\usepackage{longtable,lscape}
\usepackage{enumitem}
\usepackage{color, colortbl}
\usepackage{wrapfig}
\usepackage{xcolor,colortbl}
\usepackage{acmart-taps}

\definecolor{Gray}{gray}{0.9}
\definecolor{Gray2}{gray}{0.8}

\newcommand{\lakmal}[1]{\textcolor{black}{#1}}
\newcommand{\aurel}[1]{\textcolor{black}{#1}}

\newcolumntype{a}{>{\columncolor{Gray}}c}
\newcolumntype{b}{>{\columncolor{white}}c}

\AtBeginDocument{%
  \providecommand\BibTeX{{%
    \normalfont B\kern-0.5em{\scshape i\kern-0.25em b}\kern-0.8em\TeX}}}

 \makeatletter
\def\@fnsymbol#1{\ensuremath{\ifcase#1\or \dagger\or 
   \mathsection\or \mathparagraph\or \|\or **\or \dagger\dagger
   \or \ddagger\ddagger \else\@ctrerr\fi}}
\makeatother

\copyrightyear{2024}
\acmYear{2024}
\setcopyright{acmlicensed}\acmConference[CHI '24]{Proceedings of the CHI
Conference on Human Factors in Computing Systems}{May 11--16,
2024}{Honolulu, HI, USA}
\acmBooktitle{Proceedings of the CHI Conference on Human Factors in
Computing Systems (CHI '24), May 11--16, 2024, Honolulu, HI, USA}
\acmDOI{10.1145/3613904.3642444}
\acmISBN{979-8-4007-0330-0/24/05}

\begin{document}

\title{Learning About Social Context from Smartphone Data: Generalization Across Countries and Daily Life Moments}

\author{Aurel Ruben M\"ader}
\authornote{co-primary, listed alphabetically}
\affiliation{\institution{Idiap Research Institute \& EPFL} \country{Switzerland}}

\author{Lakmal Meegahapola}
\authornotemark[1]
\authornote{now at ETH Zurich, Switzerland}
\affiliation{\institution{Idiap Research Institute \& EPFL} \country{Switzerland}}

\author{Daniel Gatica-Perez}
\affiliation{\institution{Idiap Research Institute \& EPFL} \country{Switzerland}}

\renewcommand{\shortauthors}{M\"ader and Meegahapola, et al.}
\renewcommand{\shorttitle}{Learning About Social Context from Smartphone Data}

\begin{abstract}

%
%
Understanding how social situations unfold in people's daily lives is relevant to designing mobile systems that can support users in their personal goals, well-being, and activities. As an alternative to questionnaires, some studies have used passively collected smartphone sensor data to infer social context (i.e., being alone or not) with machine learning models. However, the few existing studies have focused on specific daily life occasions and limited geographic cohorts in one or two countries. This limits the understanding of how inference models work in terms of generalization to everyday life occasions and multiple countries. In this paper, we used a novel, large-scale, and multimodal smartphone sensing dataset with over 216K self-reports collected from 581 young adults in five countries (Mongolia, Italy, Denmark, UK, Paraguay), first to understand whether social context inference is feasible with sensor data, and then, to know how behavioral and country-level diversity affects inferences. We found that several sensors are informative of social context, that partially personalized multi-country models (trained and tested with data from all countries) and country-specific models (trained and tested within countries) can achieve similar performance above 90\% AUC, and that models do not generalize well to unseen countries regardless of geographic proximity. These findings confirm the importance of the diversity of mobile data, to better understand social context inference models in different countries.

\end{abstract}



\maketitle

\section{Introduction}

\lakmal{Human beings inherently need social interactions, community, and interpersonal relationships \cite{baumeister1995need}. As studies have shown, a fulfilling social life is particularly important for the mental well-being of young adults \cite{fuligni1993perceived, weeks2012loneliness, blakemore2012development}. Regular interaction with friends, family, or colleagues during adolescence positively impacts long-term mental and physical health. Conversely, social isolation, such as living alone or prolonged solitude, increases the likelihood of depression and other mental and physical health issues \cite{holt2015loneliness, matthews2016social, vanhalst2015lonely, baumeister1995need, blakemore2012development}. Therefore, understanding an individual's social context over time is beneficial in mobile health applications, enabling timely interventions and feedback \cite{meegahapola2020alone, meegahapola2021examining}. Beyond that, "a system is context-aware if it uses context to provide relevant information and/or services to the user, where relevancy depends on the user’s task" \cite{dey2001understanding}, and social context is a key attribute in making mobile applications context-aware, enhancing their relevance and effectiveness \cite{raento2005contextphone, liang2015social, chen2000survey, dey2001conceptual, porfirio2020transforming, mayer2015making}. Hence, inferring social context could also benefit context-aware computing, enabling devices and applications, like smartphones, to adapt functionalities (i.e., displaying notifications, silencing the phone, recommendations, etc.) based on the user's social setting. Hence, this inference not only improves user engagement through personalized interactions, but also aids digital health applications to provide timely support and enhance the overall well-being of users.}


\lakmal{While there is no single definition of social context, the notion of whether a person is alone or not has been a key construct in previous studies \cite{burns2011harnessing, meegahapola2021examining, kammoun2023understanding, meegahapola2020alone, wang2023detecting}. Social context, similar to mood, location, and food consumption, has traditionally been studied using surveys and ecological momentary assessments \cite{meegahapola2021examining, ben2015next}. However, these techniques often lead to user burden and sparse data. As an alternative, mobile sensing, which involves using smartphone and wearable sensor data-based inferences, offers a passive method for data collection \cite{meegahapola2020smartphone}. This approach is a part of the broader fields of ubiquitous computing and human-computer interaction, focusing on passively sensing a wide range of individual contexts \cite{radu2018multimodal, consolvo2008activity, wang2020social, nepal2022covid, yurur2014context, devlic2009context, adams2008sensing}. Nevertheless, accurately determining whether a person is alone or not with smartphone sensing is still challenging \cite{ali2011social, yan2013smartphone, burns2011harnessing, yurur2014context, meegahapola2020smartphone, meegahapola2020alone, hu2017elderly, solmaz2017together}. Previous studies in this direction have generally focused on specific situations, such as during eating or drinking events, or within specific populations like those with depression \cite{meegahapola2020alone, meegahapola2020smartphone, burns2011harnessing}. This leaves a research gap in understanding how social context interplays with various behavioral and contextual factors, including mood, location, time of day, and activities. Moreover, the effectiveness of smartphone sensor-based models in inferring social context in daily life has yet to be fully explored.}

\lakmal{Data diversification in machine learning involves using a heterogeneous dataset for training models to enhance their representativeness and ability to generalize across different populations \cite{gong2019diversity}. This concept, vital in domains like computer vision \cite{buolamwini2018gender, shankar2017no} and natural language processing \cite{wagner2015s, bolukbasi2016man}, focuses on the country of origin of data as a crucial diversity factor. Such diversity not only enhances model robustness, but also ensures fairness in the outcomes, benefiting a wide range of users globally \cite{dwork2012fairness, zou2018ai}. However, applying data diversification to mobile sensing has been challenging in research, due to the scarcity of large-scale datasets from various countries, collected with similar protocols \cite{meegahapola2023generalization, assimeegahapola2023complex}. The large cost and effort required to gather phone sensor data across multiple countries have been major obstacles. Hence, only a few studies have examined the effect of country diversity in mobile sensing-based models \cite{khwaja2019modeling, meegahapola2023generalization, nanchen2023keep}, and notably, none have focused on inferring social context \cite{phan2022mobile}. Hence, we ask the following research questions.} 

\begin{itemize}[wide, labelwidth=!, labelindent=0pt]

    \item \textbf{RQ1:} Which situational and behavioral aspects are associated with different social contexts in a five-country dataset?
    
    \item \textbf{RQ2:} Can the social context of daily life moments be inferred using smartphone sensing with a non-personalized, one-size-fits-all, multi-country approach? Does the separation of sensing data by its origin countries aid the performance?
    
    \item \textbf{RQ3:} What is the effect of training partially personalized models on model performance?
    
    \item \textbf{RQ4:} Do social context inference models generalize well to unseen countries?
    
\end{itemize}
\vspace{-0.05 in}

\lakmal{In answering the above RQs, in line with terms provided in Table~\ref{tab:terminology}, our work provides the following contributions:}

\vspace{-0.05 in}

\begin{itemize}[wide, labelwidth=!, labelindent=0pt]
    \item \textbf{Contribution 1: } \lakmal{We analyzed a large smartphone sensing dataset collected from 580 young adults across five countries: Denmark, Italy, Mongolia, Paraguay, and the UK. This dataset comprises passive sensing data from various modalities (Table~\ref{tab:sensor_features}) and over 216K social context self-reports gathered over a period of four weeks. A statistical analysis revealed that features from modalities such as the amount of app usage (i.e., tools, communication, productivity, social, etc.), time spent doing activities (i.e., in a vehicle, on foot, walking, etc.), step count, and screen episodes were among the top ten in terms of statistical significance for discerning alone versus not alone, based on mixed effects models. Furthermore, we observed that these features vary across countries, highlighting the diversity in individual behaviors and the corresponding contextual factors in different social contexts.}
    
    \item \textbf{Contribution 2: } \lakmal{Our study compared three approaches to social context inference: country-specific, country-agnostic, and multi-country. The multi-country approach yielded a moderate AUC (67.61\% without feature selection, 68.0\% with it), while country-specific models showed similar performance, generally within a 5\% performance difference. Notably, Paraguay and Mongolia exceeded the multi-country model's performance, unlike the other three countries. This finding challenges the notion common in previous multi-country sensing studies that country-specific models are always superior, suggesting that for social context inference, this may not necessarily be the case.}

    \item \textbf{Contribution 3: } \aurel{We found that hybrid models yielded similar performance for both multi-country and country-specific approaches. While the multi-country model achieved an AUC of 95.3\%, countries such as the UK, Denmark, and Mongolia slightly outperformed this performance, whereas Italy and Paraguay were marginally below it. These results show that with personalization, social context in everyday life could be inferred with high performance.}

    \item \textbf{Contribution 4: } \aurel{In the specific cases of mood inference and complex daily activity recognition, even though prior work found that models might generalize reasonably to geographically closer countries in Europe, we did not find such associations in social context inference performance. In fact, in all cases, models did not generalize well to unseen countries regardless of the geographical proximity. This raises concerns regarding where a trained social context inference models could be deployed.}  
\end{itemize}



\begin{table*}
  \caption{Terminology and definitions used in the paper.}
  \Description{This table provides terminology and definitions used in this paper, including terms like social context, country-specific approach, country-agnostic approach, multi-country approach, population-level model, and hybrid model.}
  \label{tab:terminology}
  \resizebox{0.8\textwidth}{!}{
  \begin{tabular}{l p{10.5cm}}
    
    \cellcolor[HTML]{EDEDED}\textbf{Term} &
    \cellcolor[HTML]{EDEDED}\textbf{Description}
    \\
    
    
    Social Context & 
    If a person is Alone or Not (i.e., with friends, family, or others). We borrow this definition from prior studies in human-computer interaction and ubiquitous computing domain \cite{burns2011harnessing, meegahapola2021examining, kammoun2023understanding, meegahapola2020alone}.
    \\

    \arrayrulecolor{Gray}
    \hline 
    
    Country-Specific Approach &
    Machine learning models are trained and tested only on data from one specific country.
    \\
    
    \arrayrulecolor{Gray}
    \hline 
    
    Country-Agnostic Approach &
    Machine learning models are trained on one country or a set of countries and tested on another country not seen in training. Such models are usually trained with the assumption that the model could be deployed to other countries, hence the term agnostic. 
    \\
    
    \arrayrulecolor{Gray}
    \hline 
    
    Multi-Country Approach & 
    Machine learning models are trained on a set of countries and tested again on the same set of countries. This is the generic way of training models when data from multiple countries are available. \\
    
    \arrayrulecolor{Gray2}
    \hline

    Population-Level Model (PLM) &
    Training and Testing splits have a disjoint set of users. Represents a case where a machine learning model trained with a population is deployed to a mobile app that a new user uses. Hence, end-user data have not been used in model training, leading to non-personalized and generic one-size-fits-all models.\\

    \arrayrulecolor{Gray}
    \hline
    
    Hybrid Model (HM) &
    Training and testing splits do not have a disjoint set of users. Represent a case when a mobile app user uses a machine learning model for some time, and data from the user is used in re-training (or fine-tuning in the case of neural networks) models. Hence, this approach leads to partially personalized models. \\

    \arrayrulecolor{Gray2}
    \hline 
    
\hline 
    
  \end{tabular}}
\end{table*}

\section{Background and Related Work}\label{section:lit}

Next, we surveyed various lines of research that our work draws upon and grouped them into three main areas: \emph{(i)} social context inference; \emph{(ii)} context recognition and health sensing; and \emph{(iii)} leveraging diversity-awareness. 

\subsection{Social Context Recognition}

The social context of individuals is a multifaceted concept encompassing complex phenomena operating on various dimensions. Two dimensions that are of particular importance are \textit{(i)} the number of people in the individual's surroundings, and \textit{(ii)} the type of relationship associated with the individual's current social context, such as family, colleagues, partner, etc. \cite{adams2008sensing, sawyer2012using, mayer2015making, bhattacharjee2023integrating}. A fundamental consideration in this context is whether an individual is alone or not, an aspect that is often discussed in mobile sensing studies exploring topics like depression, mood, eating behavior, drinking behavior, and overall mental well-being \cite{burns2011harnessing, servia2017mobile, berke2011objective, meegahapola2021examining, meegahapola2020alone}. Hence, as mentioned in Table~\ref{tab:terminology}, we use alone or not alone as a fundamental construct for social context\footnote{\lakmal{In this study, we chose the term "social context" instead of other possible terms such as "social encounters" or "social interactions". This decision was guided by the ambiguity inherent in these alternatives. "Social encounter" and "social interaction" imply a degree of engagement that we could not definitively ascertain given the study design \cite{heron1970phenomenology, hu2017elderly}. Our term aligns with previous ubicomp and human-computer interaction studies \cite{meegahapola2020alone, meegahapola2021examining, kammoun2023understanding}, which often use "social context" to differentiate between being alone versus not alone. We acknowledge this as a limitation and recognize that our use of "social context" carries a narrower definition specific to our study's design.}}.


Previous studies examining the task of social context sensing have frequently delved into ethical and technical considerations \cite{ali2011social} or have aimed to contextualize the task within the \lakmal{broader scope} of social networks and user behavior \cite{adams2008sensing, wang2023detecting}. An intriguing paradox related to the internet, particularly relevant in such analyses, is the way mobile phones serve as platforms to replicate and, to some extent, replace physical social contexts. This phenomenon is particularly notable among young adults who increasingly substitute face-to-face interactions with mobile-mediated communication \cite{twenge2019less}. It could be argued that the "connection technology" of the Internet, in many instances, has heightened social disconnection \cite{adams2008sensing}. \lakmal{Another recent study \cite{wang2023detecting} examined the social context, which they referred to as being alone vs. dyadic vs. in a group, during virtual conversations of a group of socially anxious undergraduate students. With a multi-task learning model, they showed that passive sensor data from Empatica E4 devices could detect contexts relevant to social anxiety. However, it is crucial to clarify that in the context of our study, the task of social context inference pertains to the presence of individuals in physical, in-person settings and does not encompass virtual connections formed through mobile phones, virtual conversations via video conferencing applications, or social networks.}

Several studies have attempted to infer in-person social context, specifically whether individuals are alone or not, utilizing sensing data. In a study by Burns et al. \cite{burns2011harnessing}, sensor data was collected from eight participants over eight weeks, involving a total of 38 different sensors. They employed regression trees to predict an individual's social context, achieving mean accuracies of 80\% across participants. It is worth noting that this study focused on a small sample of individuals diagnosed with depressive symptoms, \lakmal{leaving questions regarding the generalizability of their findings to a broader population without such conditions}. Other studies have concentrated on social context inference within specific contextual situations. For instance, in a study by Meegahapola et al. \cite{meegahapola2020alone}, the authors explored the social context of university students during eating episodes, emphasizing the significance of determining whether an individual is alone or not for providing timely interventions and feedback in mobile food diaries. They trained and evaluated their models with data collected during eating events. Further, they used population-level models and achieved accuracies ranging from 70\% to 75\% for Switzerland and Mexico, separately. However, a limitation of this study is that direct comparisons between the results of the two countries are challenging due to variations in data collection methods, sensor usage, time periods, and protocols. Additionally, their sample sizes were relatively smaller compared to the current study. In a related work by Meegahapola et al. \cite{meegahapola2021examining}, the social context of young adults during alcohol drinking episodes was examined. Using population-level models, they attained accuracies ranging from 80\% to 87\% for various two-class social context inferences, including alone or not, as well as alone or with friends/family/partners. However, this study was based on a dataset collected in a single European country and exclusively revolved around the social context of drinking events. Furthermore, while both of these studies involved healthy participants without eating, drinking, or mental well-being-related disorders, they were primarily centered on specific contextual scenarios like eating and drinking. Therefore, the social context of young adults in daily life moments has remained relatively unexplored in prior research employing smartphone sensing data. Additionally, prior work has not delved deeply into analyzing country-level diversity and model personalization approaches, which are also focal points of our study.

\subsection{Context Recognition and Health Sensing}

\lakmal{In the mobile sensing literature, social context inference can be seen as a context recognition task. Following the framework that categorizes mobile sensing-based inferences into three pillars: behavior, person, and context, social context inference falls within the third pillar \cite{meegahapola2020smartphone}. Furthermore, context awareness in devices offers significant benefits, such as battery life optimization. For example, reducing sensor sampling rates like GPS and cellular sensors during low battery usage periods, like sleep or social events, can conserve battery \cite{yurur2014context}. Devices can also adapt behaviors to specific situations, like disabling calls or switching to silent mode during meetings \cite{lockhart2012applications}. Additionally, mobile sensing has been extensively used to detect health-related contexts, including stress, mood, depression, energy expenditure, and eating and drinking behaviors, along with their social implications \cite{can2019stress, likamwa2013moodscope, burns2011harnessing, bae2017detecting, meegahapola2020alone, amarasinghe2023multimodal}. These studies aim to identify, analyze, and intervene in health-related situations. Accurate context identification is crucial for further analysis or intervention \cite{berke2011objective}. Once a health-related context is identified, it enables the exploration of associated factors, like peer pressure in heavy drinking episodes among college students \cite{meegahapola2021examining}, or depression linked to prolonged solitude \cite{matthews2016social}. Applications developed for health interventions, such as those for depression \cite{burns2011harnessing, wahle2016mobile} or mood tracking \cite{servia2017mobile}, are based on accurately identifying these contexts. Hence, social context recognition, given its direct impact on mental and physical well-being, is a crucial subset of health sensing \cite{holt2015loneliness, matthews2016social}. However, social context recognition in broad daily life situations, from smartphone data, has rarely been explored}.

\subsection{Leveraging Diversity Awareness}

Systems that rely on learning patterns from data, such as classification and regression models, may inadvertently encode the internal biases present in the training data \cite{zou2018ai}. Consequently, these biases can manifest in the form of biased predictions, particularly concerning gender, race, socio-economic status, or the psychological profile of new users, when applying pre-trained models. In the realm of computer vision research, the impact of biased training data has been extensively examined. For example, one study evaluated gender classification models across various subgroups defined by gender and race \cite{buolamwini2018gender}. This study revealed considerable disparities in model performance, with dark-skinned women experiencing error rates as high as 34.7\%, while error rates for light-skinned men were only around 0.8\%. These biases frequently stem from the under-representation of certain groups within the datasets used for model training.

\lakmal{Popular image classification datasets like ImageNet \cite{deng2009imagenet} and Open Images \cite{krasin2017openimages} have faced criticism for under-representing certain regions, with only a small fraction of images from countries like China and India \cite{shankar2017no}. This leads to biased predictions when models trained on these datasets are used in these countries \cite{shankar2017no}. In natural language processing, similar biases are evident, such as gender stereotypes in word embeddings and gender-biased translations in software like Google Translate \cite{bolukbasi2016man, prates2020assessing}. Addressing these imbalances is crucial for improving both the performance and fairness of machine learning systems \cite{zou2018ai}. In mobile sensing, diversity across datasets poses unique challenges, with biases not as apparent as in other fields. Strategies like partitioning datasets into smaller subsets \cite{abdullah2012towards} and considering wearables' usage diversity \cite{chang2020systematic} have been proposed, but the impact of geographical diversity on sensor-based tasks remains underexplored \cite{phan2022mobile}, underscoring the need for thorough investigation in this area.}

\lakmal{To enhance diversity in mobile sensing datasets, collecting data from diverse geographic regions is a key \cite{schelenz2021theory}. This approach, however, has seen limited exploration in mobile sensing research. For example, Khwaja et al. \cite{khwaja2019modeling} analyzed data from five countries (the UK, Spain, Colombia, Peru, and Chile) for personality trait inference, finding that country-specific models were more effective than multi-country models, with performance improvements ranging from 3\% to 7\%. Similar benefits of country-specific models were observed in mood inference and complex daily activity recognition \cite{meegahapola2023generalization, assimeegahapola2023complex}. However, the applicability of these findings to social context inference remains to be further examined.}

\lakmal{In summary, this study introduces several novel aspects to the field of social context inference using smartphone sensing data. Firstly, it expands on previous research by identifying key smartphone sensing features and behavioral and contextual factors from self-reports that are instrumental in inferring social context, uniquely extending the analysis to multiple countries. This is conducted using an extensive, multimodal smartphone sensing dataset, which sets our study apart in terms of model generalization to everyday life situations across different countries. Secondly, we explore the feasibility of social context inference using a generic multi-country approach, employing both population-level and hybrid models. This represents one of the earliest efforts in social context inference with smartphone data from a considerable number of participants. Thirdly, our study includes a comprehensive comparative analysis using frameworks from previous research, examining country-specific, country-agnostic, and multi-country approaches. This analysis shows the effectiveness of multi-country models in social context inference. However, it also reveals the challenges in generalizing these models to unobserved countries, underscoring the complexity of social context inference in diverse settings.}

\section{Dataset}\label{sec:dataset}

\begin{table*}
  \caption{Dataset Summary.}
  \Description{Summary statistics of the dataset used in this study are presented, covering five countries: Denmark, Italy, Mongolia, Paraguay, and the UK. The paper uses data from 581 participants who provided over 292,000 self-reports on their social context, with a roughly equal distribution between 'alone' (around 51\%) and 'not alone' cases.}
  \label{tab:summary_table}
  \begin{tabular}{l p{2.4cm} p{6cm} r r}

    \cellcolor[HTML]{EDEDED}\textbf{Country} & 
    \cellcolor[HTML]{EDEDED}\textbf{\# of Participants} &
    \cellcolor[HTML]{EDEDED}\textbf{\# of Participants With Sufficient Data} &
    \cellcolor[HTML]{EDEDED}\textbf{\# of Reports} &
    \cellcolor[HTML]{EDEDED}\textbf{Alone Percentage} 
    \\
    
    UK & 72 & 53 &  26,687 & 69.05\% \\
    Denmark & 25 & 17 & 10,058 & 49.63\% \\
    Italy & 238 & 221 & 151,335 & 64.78\% \\
    Paraguay & 29 & 24 & 9,745 & 54.52\% \\
    Mongolia & 217 & 138 & 94,249 &  24.52\% \\
    
    \arrayrulecolor{Gray}
    \midrule
    
    All &  581 & 453 & 292,074 & 51.31\% \\
    
    \arrayrulecolor{Gray2}
    \bottomrule 
  \end{tabular}
\end{table*}

For this study, we used a novel smartphone sensing dataset from our previous work, that was collected in a study conducted simultaneously in five countries \cite{assimeegahapola2023complex, giunchiglia2022worldwide}. To the best of our knowledge, this is the only dataset that allows us to perform the analysis around social context during daily life moments, across multiple countries, with smartphone data. The data were collected for four weeks from college students of the following five universities: Aalborg University (Denmark), London School of Economics and Political Science (the United Kingdom), the National University of Mongolia (Mongolia), the Universidad Católica "Nuestra Señora de la Asunción" (Paraguay), and the University of Trento (Italy). The study participants contributed three different kinds of data: \emph{(i)} closed-ended questionnaires, \emph{(ii)} hourly self-reports throughout the day, and \emph{(iii)} sensor data.

\emph{(i)} The closed-ended questionnaires consisted of three separate questionnaires, which were administered to the study participants before the start of the study, before the start of the sensor data collection, and after two weeks of sensor data collection. This design allowed the collection of a large amount of information from participants without overburdening them. The questionnaires were designed to capture surface diversity information (age, sex, country, etc.) and deep diversity information (personality, values, intelligence, etc.--- with validated scales).

\emph{(ii)} During the data collection, a mobile app was deployed. Using the app sensor data were captured passively, and participants also self-reported details about their behavior and context. These hourly self-reports were meant to capture how people spend their time. A notification was sent to participants once every hour, and they were asked to report their current activities (studying, cooking, etc.), semantic location (home, library, etc.), mood (valence--- in a five-point scale from very negative to very position), and social context (alone, friends, relatives, classmates, roommates, colleagues, partner, other). Hence, the dependent variable for the discussed alone-or-not inference is based on the answers from the social context self-report.

\emph{(iii)} The sensor data originally consisted of 34 different sensors, which are divided into continuous and interaction sensing. Continuous sensing modalities (captures sensor data regardless of user activities) included activity type, step count, Bluetooth, WiFi, location, cellular, and proximity; and interaction sensing modalities (measures the interactions participants have with the phone) included app usage, touch events, the screen on/off episodes, notification, etc. 

In total, four weeks of sensor and self-report data were collected from study participants. However, our analysis revealed that the completeness and the number of responses to time diaries and questionnaires vary greatly between different countries (Table \ref{tab:summary_table}) and different participants (Figure \ref{fig:events_per_part}). Furthermore, the quality and quantity of sensor data also vary between different participants. This is also observable in the amount of missing sensor data (Nan) for certain sensors or study participants (Figure \ref{fig:nan_ratio_sensors}). A more detailed discussion about the dataset and feature extraction can be found in papers by original authors that presented the dataset \cite{meegahapola2023generalization, assimeegahapola2023complex}.

The final analysis of this study consists of inferring the social context of a participant. Given the large size of the original dataset, which amounts to approximately 30GB of data in different data formats, the sensor data were processed and aggregated to sensible features at the self-report level. To do this, we used an approach similar to prior mobile sensing studies where sensor data are aggregated around the time of an in-situ self-report \cite{servia2017mobile, bae2017detecting, biel2018bites, meegahapola2021examining, meegahapola2020alone}. More specifically, all sensor data corresponding to ten minutes around the social context self-report was aggregated to features as summarized in Table~\ref{tab:sensor_features}. Thus, the aggregation time frame corresponds to a ten-minute window around the answering time of a self-report, which captures the individual's current activity, semantic location, mood, and, finally, the social context. \lakmal{It is worth noting that various time windows ranging from 1 to 20 minutes were explored. Extending beyond a 20-minute window was not feasible due to the risk of data leakage from overlapping sensor data segments. Among the tested durations, the ten-minute window emerged as reasonable, offering satisfactory performance. Additionally, this specific time frame aligns with previous studies that used the same dataset \cite{meegahapola2023generalization, assimeegahapola2023complex}, thereby facilitating meaningful comparisons.} Moreover, in total, 117 distinct sensor features were created in line with the work of the original authors \cite{assimeegahapola2023complex}, including 48 features that correspond to app usage time period based on app categories in Google Playstore (e.g., action, dating, music, puzzle, social, etc.) similar to \cite{santani2018drinksense, likamwa2013moodscope}. Additionally, time and day features were also included in the analysis, where the day of the week corresponds to a categorical variable, weekday or weekend, while the time was noted with the hour of the corresponding time dairy event (a numeric value between 0 and 23). Furthermore, missing data markers were introduced, which indicate when data is missing for certain sensor groups \cite{little2019statistical}. In summary, the smartphone sensor features of the final dataset used for inference are detailed in Table~\ref{tab:sensor_features} in the Appendix.

\aptLtoX[graphic=no,type=html]{\begin{figure}
        \centering
        \includegraphics[width=0.8\linewidth]{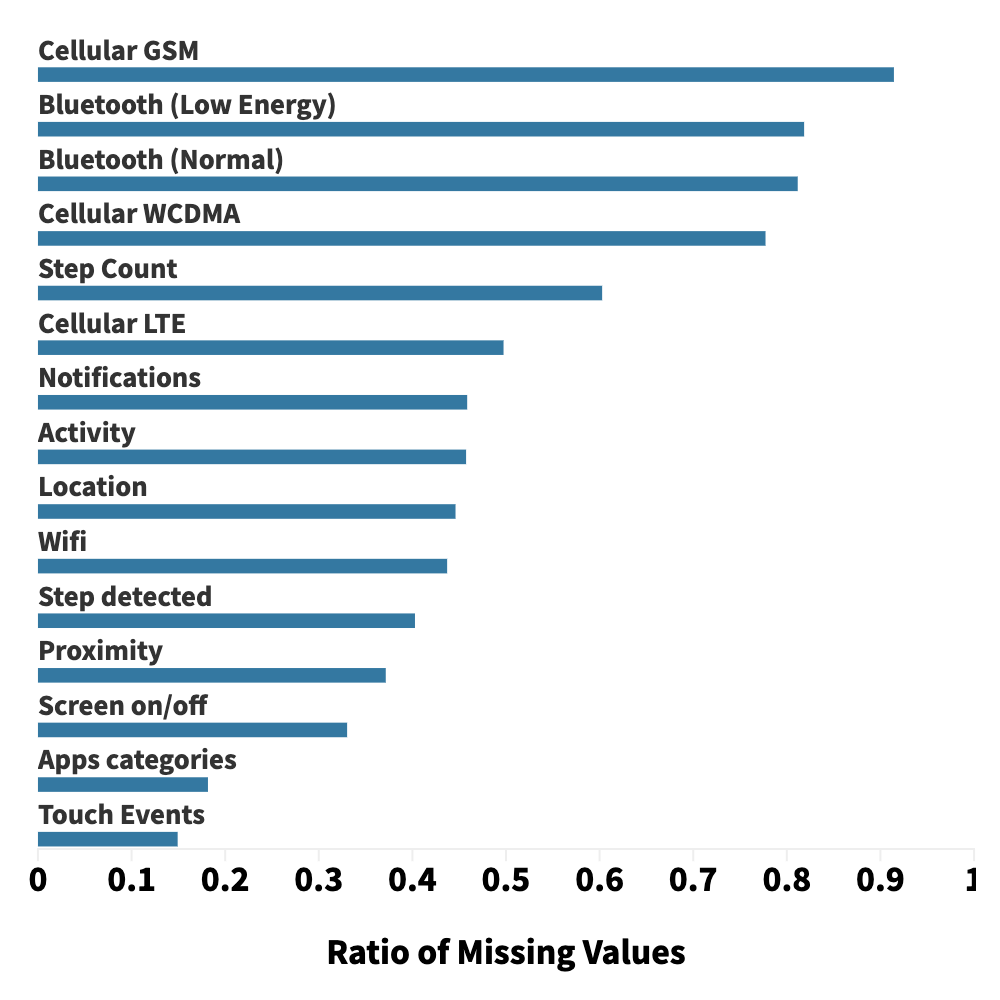}
        \caption{Ratio of missing data (Nan values) in percentage aggregated to sensor group.}
        \Description{This figure illustrates the percentage of missing data (NaN values) aggregated by the sensor group. It shows that cellular signals have the highest missing values, while touch events have the lowest.}
        \label{fig:nan_ratio_sensors}
    \end{figure}
    \begin{figure}
        \includegraphics[width=0.8\linewidth]{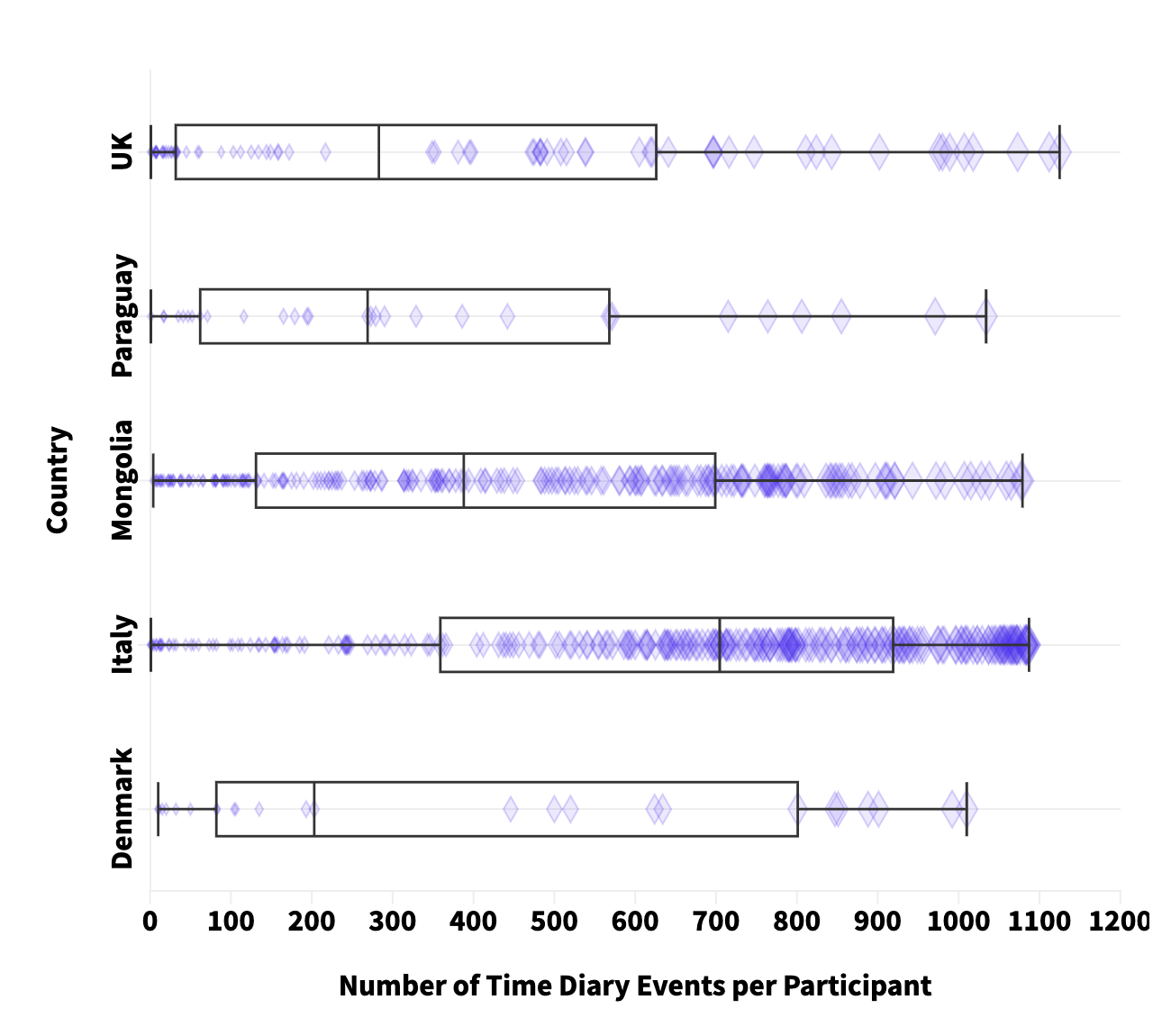}
        \caption{Boxplot of the number of self-reports per participant, grouped by country. The boxes extend till the upper and lower quartiles, the median is indicated by an orange line, and whiskers extend till 1.5x IQR.}
        \Description{A boxplot representing the number of self-reports per user, categorized by country. The boxes indicate the upper and lower quartiles, with an orange line for the median and whiskers extending to 1.5 times the interquartile range. Italy generally has the highest per-person self-reports, while Denmark has the lowest.}
        \label{fig:events_per_part}
    \end{figure}}{\begin{figure*}
    \centering
    \begin{minipage}{0.49\textwidth}
        \centering
        \includegraphics[width=0.8\linewidth]{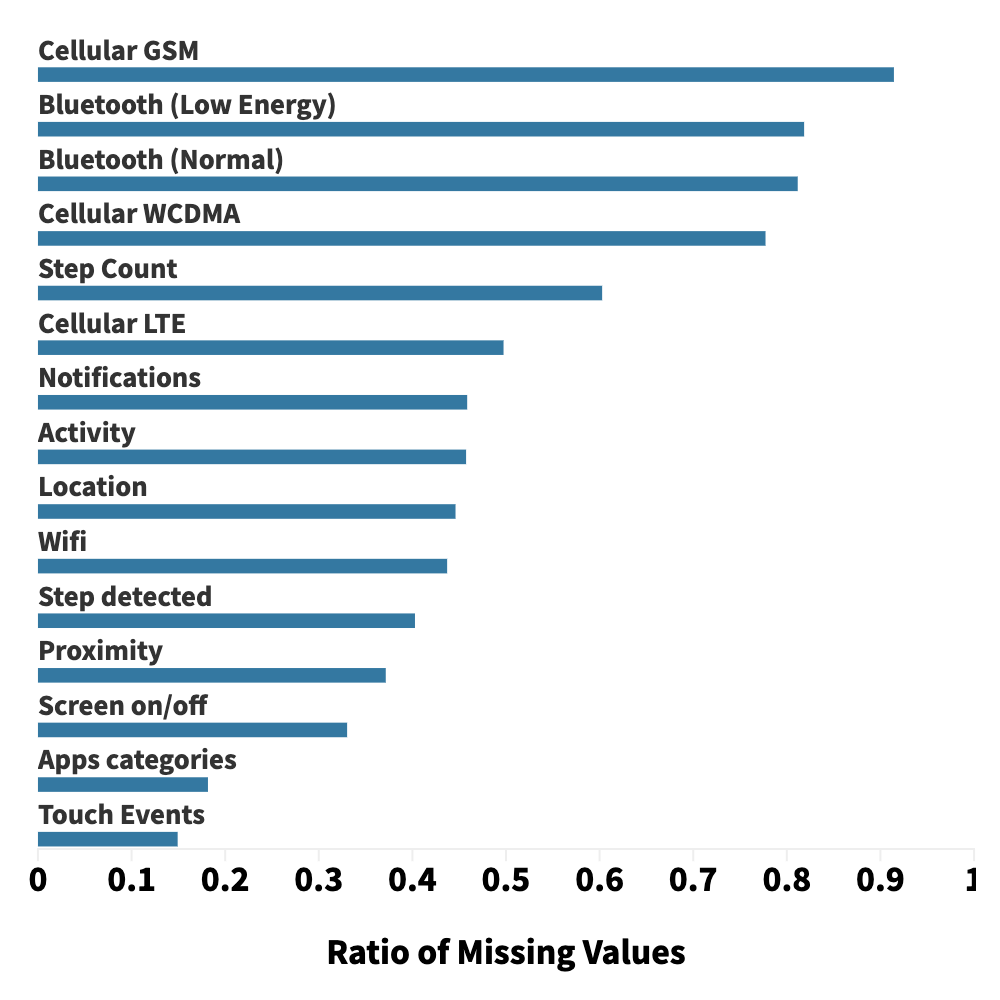}
        \caption{Ratio of missing data (Nan values) in percentage aggregated to sensor group.}
        \Description{This figure illustrates the percentage of missing data (NaN values) aggregated by the sensor group. It shows that cellular signals have the highest missing values, while touch events have the lowest.}
        \label{fig:nan_ratio_sensors}
    \end{minipage}
    \hfill
    \begin{minipage}{0.49\textwidth}
        \centering
        \includegraphics[width=0.8\linewidth]{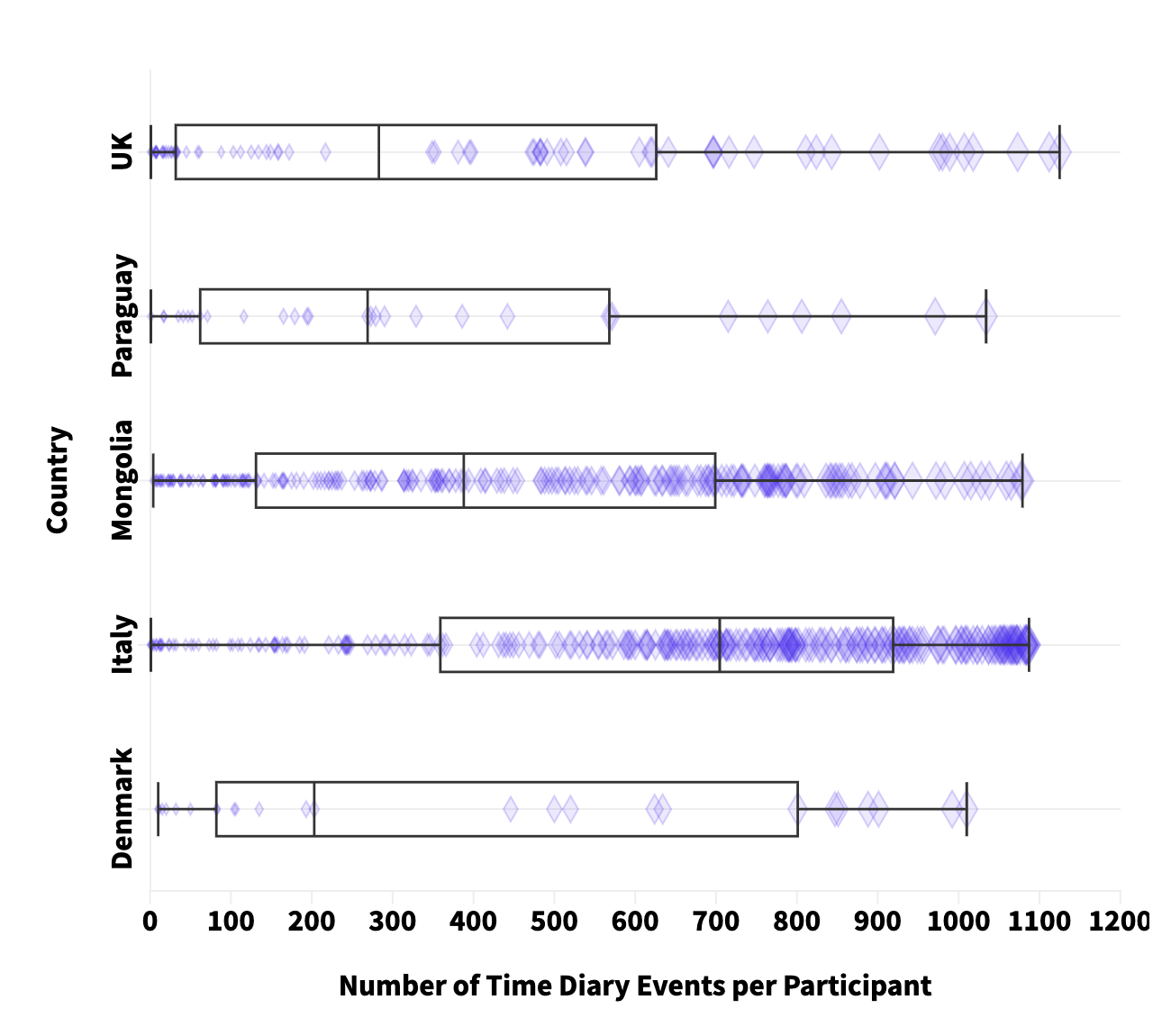}
        \caption{Boxplot of the number of self-reports per participant, grouped by country. The boxes extend till the upper and lower quartiles, the median is indicated by an orange line, and whiskers extend till 1.5x IQR.}
        \Description{A boxplot representing the number of self-reports per user, categorized by country. The boxes indicate the upper and lower quartiles, with an orange line for the median and whiskers extending to 1.5 times the interquartile range. Italy generally has the highest per-person self-reports, while Denmark has the lowest.}
        \label{fig:events_per_part}
    \end{minipage}

\end{figure*}}

\section{Experimental Setup}\label{sec:experimentalsetup}

\subsection{\textbf{RQ1: Situational and Contextual Aspects Around Social Contexts}}

The purpose of this analysis was to investigate various situational and behavioral cues related to social context (alone or not) that can be extracted from self-reports and sensor data. This examination provides us with a comprehensive overview of the dataset and positions our study in relation to prior research. Initially, the original dataset featured detailed social context categories, including labels such as alone, with classmates, with colleagues, with a partner, with relatives, with roommates, and with others. Utilizing these labels, we examined how patterns of fine-grained social contexts differed across countries (Figure \ref{fig:social_context_countries}). Subsequently, by aggregating data from all countries, we explored how social context varied across these fine-grained categories throughout the day (Figure \ref{fig:social_context_over_day}). Next, as previous studies have linked social context to mood (discussed in Section \ref{section:lit}), we conducted an analysis focusing on binary social context (i.e., alone or not) across different levels of valence, assessed using a five-point Likert scale ranging from very positive to very negative. Lastly, we delved into the variations in social context across different locations and activities (Figure \ref{fig:alone_vs_not_alone_contexts}).

\aurel{To identify which sensor features, including the hour of the day and the day of the week, are most indicative of whether someone is alone or not, we used mixed effects models \cite{sheiner1991introduction}. The results are presented in Table \ref{tab:mixed_effects_model} sorted by the descending order of statistical significance (p-values). This is done, ensuring that the most robust and least likely chance-based findings are presented first. This approach also aids in swiftly identifying the most relevant features impacting the dependent variable, making it easier to focus on key variables for further investigation or discussion. Furthermore, these experiments were conducted separately for each of the five countries, comparing the data distributions of sensing features when a participant is alone and when they are not alone. Participants were chosen as the random effect, to make sure that in-person variations are taken into account when calculating p-values. To account for multiple comparisons, we adjusted the p-values using the Bonferroni correction \cite{vickerstaff2019methods}}.

\subsection{\textbf{RQ2, RQ3 \& RQ4: Social Context Inference}}

To gain a deeper insight into the importance of country diversity in social context inference, we adopted the strategy of dividing our dataset into several subsets \cite{abdullah2012towards}, each corresponding to the origin of the data \cite{khwaja2019modeling, meegahapola2020alone}. Given that countries can exhibit distinct cultural and socio-economic norms, we presume that these factors influence people's behavior, smartphone usage, and social interactions \cite{khwaja2019modeling}. Thus, our fundamental assumption posits that data originating from a specific country is better suited for training and constructing a model specific to that country. To test this assumption rigorously, we needed to compare the results of models trained on different data splits. Therefore, drawing inspiration from prior research that has explored multi-country data \cite{khwaja2019modeling, meegahapola2023generalization, assimeegahapola2023complex}, we used several experimental approaches: \textit{i)} Multi-Country: training and testing with all available countries; \textit{ii)} Country-Specific: training and testing within the same country; and \textit{iii)} Country-Agnostic: training on one or more countries and testing on an unseen country. Indeed, these scenarios closely mirror real-world practices, where companies often endeavor to gather smartphone and wearable sensor data from diverse countries to train their models. However, there are instances where collecting data from multiple countries is impractical. In such cases, models are trained within one country or using data from a select few countries and subsequently deployed to other countries that were not included in the model's training dataset.

Additionally, we trained various machine learning models, including logistic regression, random forest classifiers, XGBoost, and AdaBoost. Initially, all models were trained using all available sensor features. Subsequently, we employed sequential forward feature selection (FS) \cite{pudil1994floating} to identify the most predictive features for training in both country-specific and multi-country approaches. Furthermore, drawing inspiration from previous research \cite{meegahapola2023generalization, meegahapola2021one}, we developed two types of models with varying levels of personalization:

\begin{itemize}
    \item Population-level models (non-personalized): Similar to leave-k-users-out strategy \footnote{\lakmal{In comparing leave-one-out and leave-k-out (k-fold, but without data from the same user across folds) evaluation methods \cite{meegahapola2021one}, we find that while leave-one-out is thorough in using almost all available data for training except for the testing user, it is computationally intensive, especially for large datasets. Leave-k-out, on the other hand, offers more flexibility and efficiency. This approach is particularly useful when dealing with larger datasets, as it balances the need for thorough evaluation with practical constraints on computational resources. Considering these factors, we chose to use the leave-k-out method for our study. This choice was driven by the need to efficiently handle our sizable dataset while still ensuring a robust evaluation of model performance.}}---that is, data in training and testing splits are not from the same users, hence $\approx$ 80:20 training and testing split. 
    \item Hybrid models (partially personalized): First, data in training and testing splits are not from the same users, and we did an $\approx$ 60:40 split. Then, add 50\% of data from testing users into the training set to achieve partial personalization for users in the testing set, leading to $\approx$ 80:20 training and testing split. \lakmal{Moreover, temporally, only the first half of data produced by testing users are removed from the testing set and included in the training set to make sure there is no temporal leakage \cite{kapoor2022leakage}. That is, no data points from the future of the testing users are used in model training.}
\end{itemize}

In summary, to address \textbf{RQ2}, encompassing \textit{(i)} the feasibility of employing a generic, non-personalized model for social context inference using multimodal smartphone sensor data, and \textit{(ii)} the potential performance improvements through data separation by countries, we implemented both multi-country and country-specific approaches with population-level models. Subsequently, to explore \textbf{RQ3}, which delves into the impact of partial personalization on model performance, we leveraged multi-country and country-specific approaches, this time with hybrid models. Lastly, to investigate \textbf{RQ4}, pertaining to the generalizability of models trained in one or more countries to unseen country data, we utilized the country-agnostic approach with population-level models in two distinct setups: Setup 1, where models were trained with data from a single country and tested on all other countries separately; and Setup 2, where models were trained with data from four countries and tested on the remaining country.

\aurel{In all cases, data were randomly sampled, and models were randomly initiated ten times with different seeds to obtain results for each setup, and results were averaged to obtain the mean and standard deviation of the area under the receiver operating characteristic curve (AUC) and F1-score metrics. To deal with the class imbalance and, for some countries, the small amount of data, Synthetic Minority Over-sampling (SMOTE) \cite{chawla2002smote} was used in training while evaluation was performed on the testing set. All results can be evaluated against a majority baseline AUC of 50\%.} Given run time restrictions, GPU implementations of the models were used. All models were run in Python with commonly used libraries such as scikit-learn \cite{pedregosa2011scikit} for logistic regression and Ada Boost, cuML \cite{raschka2020machine} as it implements a random forest model on the GPU, and lastly, a python XGBoost implementation \cite{Chen:2016:XST:2939672.2939785}. Finally, in all cases, the hyper-parameter search was done using grid search (e.g., random forest: number of trees, max depth, min sample split; XGboost: learning rate, min split loss, max depth, reg lambda, etc.).

Furthermore, behavioral sensor data often exhibit sparsity and tend to contain high volumes of missing data \cite{xu2021understanding}. Within this dataset, missing data can be attributed to various distinct scenarios: \emph{(i)} sensor malfunction, \emph{(ii)} sensor damage or absence, and \emph{(iii)} intentional sensor deactivation by the user (e.g., disabling Bluetooth/WiFi, activating flight mode, etc.). However, distinguishing between these cases based solely on the available sensor data is challenging, and as in previous studies \cite{pires2016identification, stisen2015smart}, all these scenarios are treated equivalently. The extent of missing data also varies depending on the sensor (Figure \ref{fig:nan_ratio_sensors}) and the country where the experiment was conducted (e.g., 60\% of Mongolian sensor data is missing). To address the issue of missing data, the first step involved discarding sensor groups with over 90\% missing data, aligning with prior work \cite{santani2018drinksense}. This exclusion specifically applied to the `cellular\_gsm' features (refer to Figure \ref{fig:nan_ratio_sensors}). Subsequently, the remaining missing observations were estimated using the k-nearest-neighbor (kNN) algorithm \cite{zhou2018missing, xu2021understanding}, with an experimentally determined optimal value of two for k. Following the procedure outlined in prior research \cite{assimeegahapola2023complex, zhou2018missing, xu2021understanding}, estimation was performed on the training set, and the learned model was then employed to impute missing data in the testing sets. To retain information that might be encoded in the missing entries of a given sensor group, binary markers were used to denote that the data had been interpolated \cite{little2019statistical}.

\section{Results}\label{sec:results}

\subsection{\textbf{RQ1: Situational and Contextual Aspects Around Social Contexts}}

As illustrated in Table~\ref{tab:summary_table}, the number of study participants and the number of self-reports varied depending on the location of the pilot study. In total, 581 study participants provided self-reports regarding their social context and also supplied good-quality sensor data. The number of self-reports submitted by each participant displayed substantial variation, contingent upon individual participants and their respective countries. Figure \ref{fig:events_per_part} visually depicts this diversity, indicating that the median count of self-reports per participant ranged from 200 (Denmark) to 800 (Italy). On average, participants contributed approximately 300 to 400 self-reports. Nevertheless, there were some users who provided well in excess of 1,000 self-reports, while others contributed close to zero self-reports despite providing sensor data over numerous days. To ensure an adequate volume of data for model personalization, individuals with fewer than 100 self-reports or fewer than six instances of the minority class were excluded from the analysis, resulting in a total of 453 participants for the subsequent analysis (Table~\ref{tab:summary_table}).

\begin{figure*}
    \begin{minipage}{\textwidth}
        \centering
        \includegraphics[width=\linewidth]{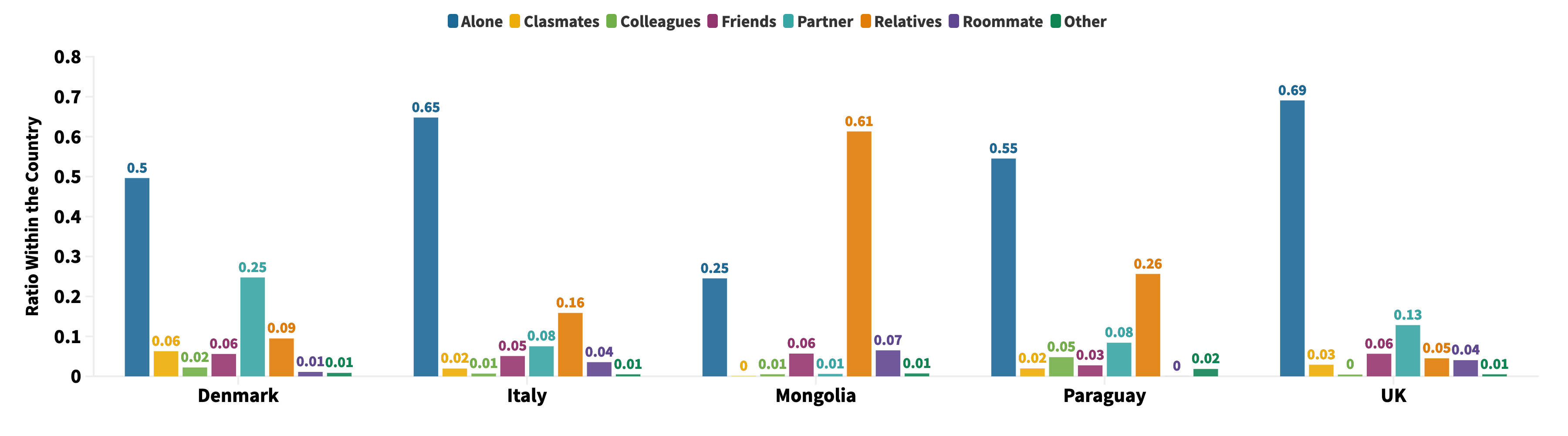}
        \caption{Distribution of social context across countries, as a ratio of the total number of self-reports.}
        \Description{This figure depicts the distribution of social contexts across countries as a ratio of total self-reports. Mongolia stands out with more 'with relatives' self-reports, in contrast to other countries where 'alone' is the dominant context.}
        \label{fig:social_context_countries}
    \end{minipage}
\end{figure*}

\begin{figure}[t]
    \centering
    \includegraphics[width=\linewidth]{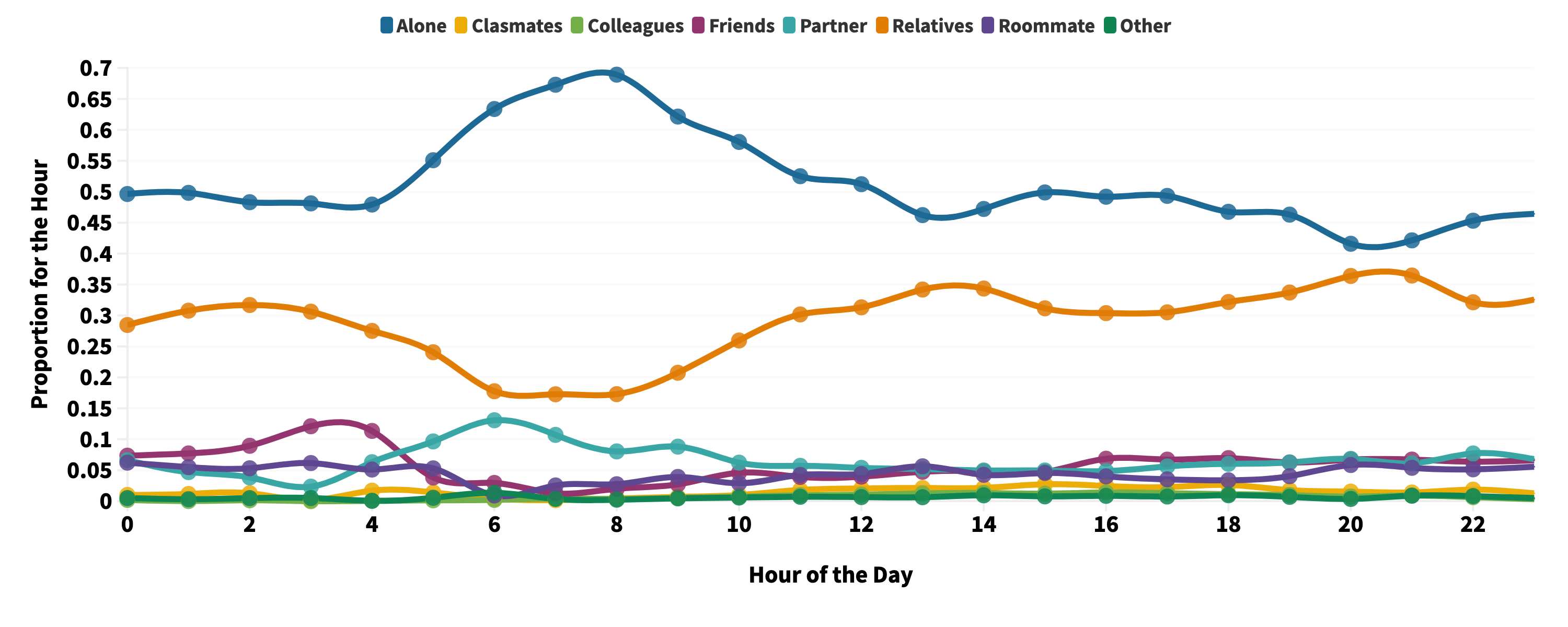}
    \caption{Social context distribution according to the hour of the day, across all participants.}
    \Description{This figure presents the distribution of social context according to the hour of the day for all participants. It shows that self-reports of being alone peak around 8 am and dip at around 8 pm, while reports of being with relatives peak at around 8 pm.}
    \label{fig:social_context_over_day}
\end{figure}

\lakmal{The distribution of social contexts exhibits differences depending on the participant's respective country, as evident in Figure \ref{fig:social_context_countries}}. \lakmal{In most countries, with the exception of Mongolia, individuals predominantly self-reported to have been alone.} However, in Mongolia, participants reported to have been in the company of their families in over 60\% of cases. This could suggest varying cultural norms related to living arrangements or family interactions. \lakmal{Although it is not possible to generalize the culture and practices of the entire Asian region to Mongolia, prior work has suggested that Asian households often have stronger family ties compared to Western societies \cite{goody1996comparing}. Consequently, young adults in these societies may tend to live with their parents for longer periods \cite{kim2015relationships, zhao2018impact, cheung2021systematic}}. It is indisputable that the participant's country influenced the prevalence of different social context types (e.g., alone, with family/friends/partner, etc.) within the dataset. For example, in Denmark and the UK, participants reported spending more time with their partners, sometimes even more than with their families. \lakmal{Country-specific differences in the frequency and nature of social contexts are corroborated by findings in life sciences literature \cite{taniguchi2021family} and recent reports \cite{ortiz2020time}.}

\begin{figure*}%
    \centering
    \subfloat[\centering Valence ]{{\includegraphics[width=7.5cm]{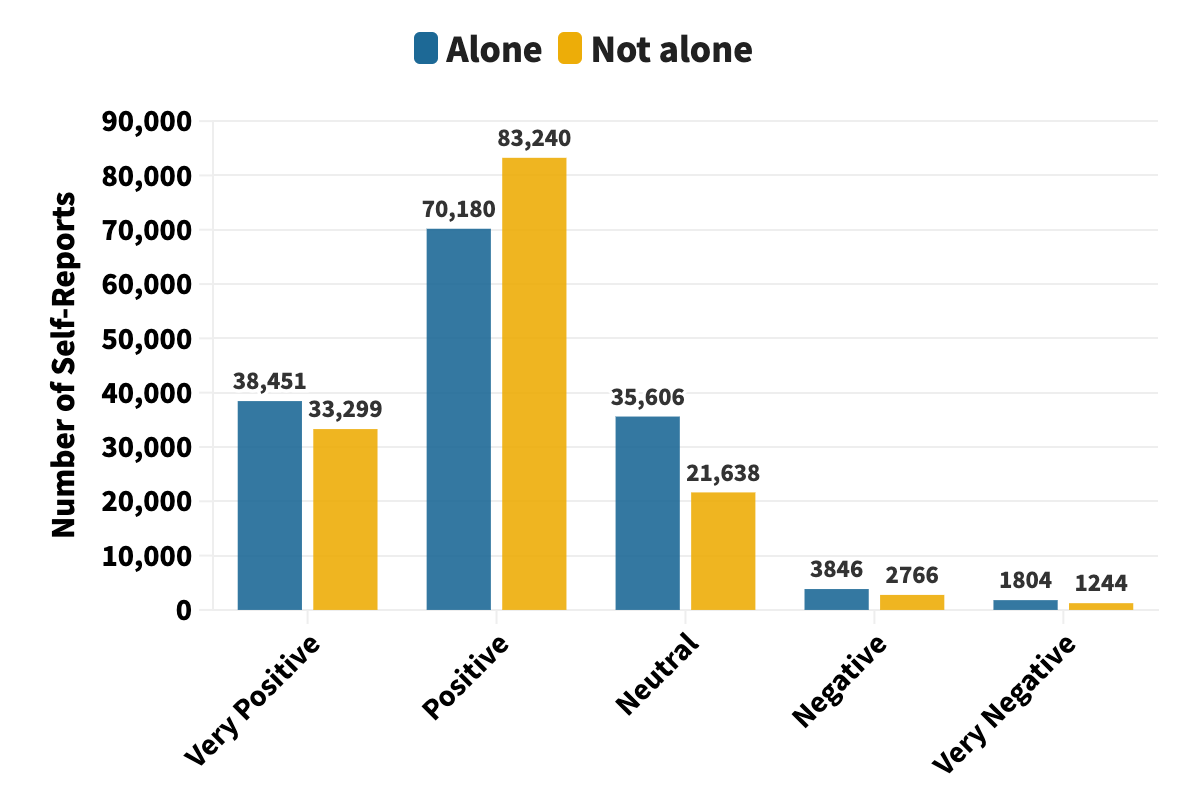} }}%
    \hfill
    \subfloat[\centering Semantic Location]{{\includegraphics[width=7.5cm]{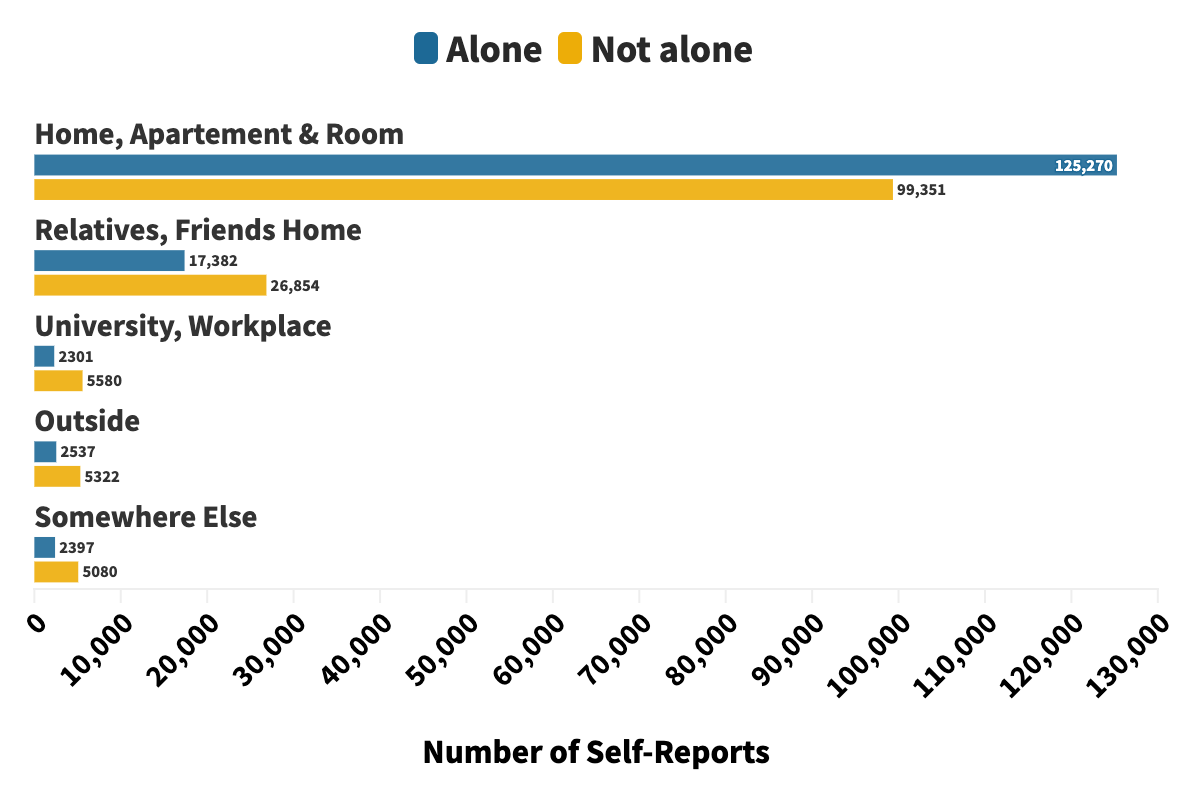} }}%
    \hfill
    \subfloat[\centering Concurrent Activity ]{{\includegraphics[width=12cm]{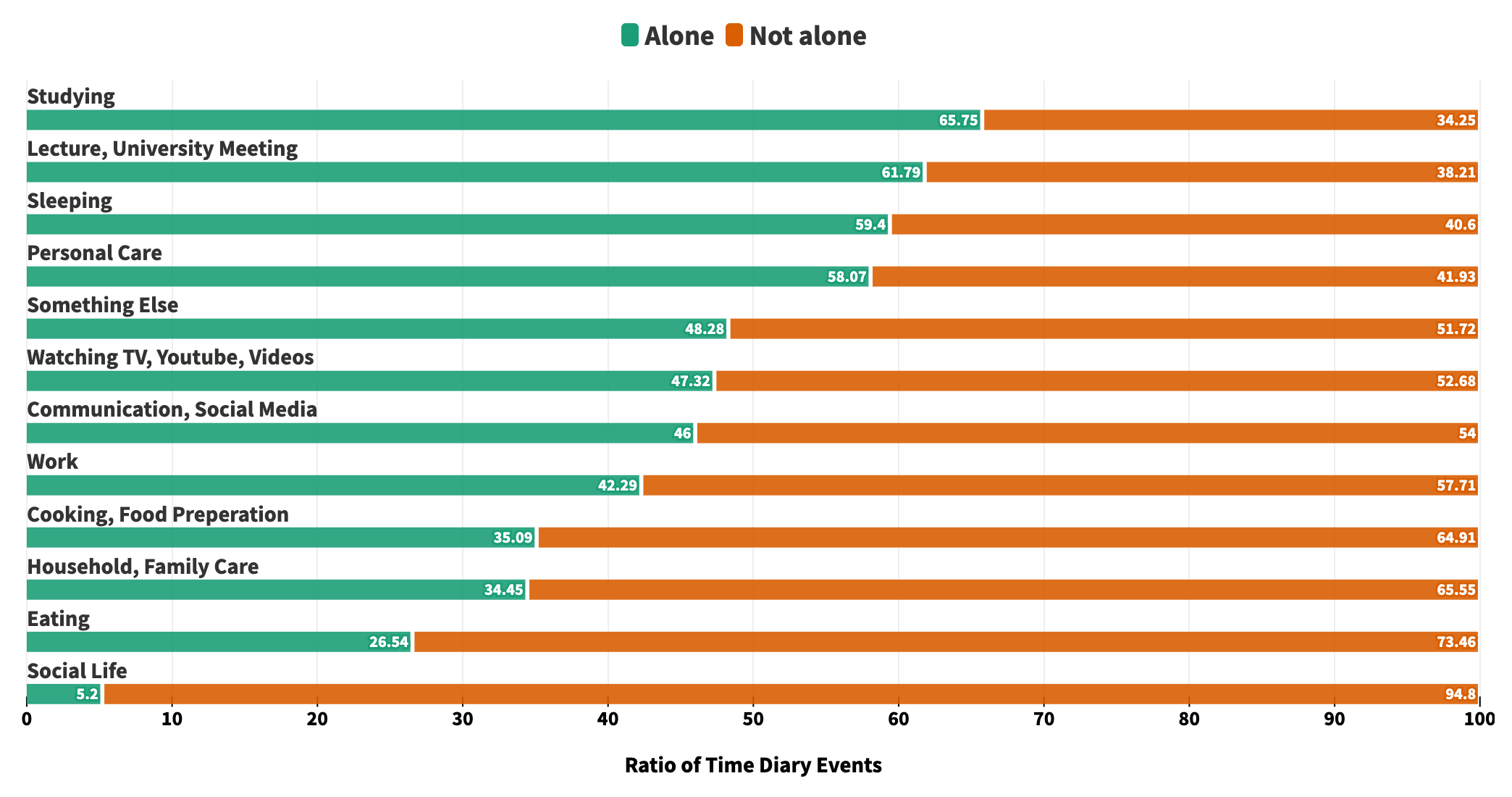} }}%
    \caption{Alone or not alone depending on a) Valence, b) Semantic Location, and c) Concurrent Activity of the participants.}%
    \Description{The figure compares 'alone' and 'not alone' self-reports based on participants' valence, semantic location, and concurrent activity. It reveals a higher proportion of 'not alone' reports associated with positive valence and more 'alone' reports in other valences. 'Alone' reports are more frequent at home, while 'not alone' reports dominate in other locations. 'not alone' reports have the highest percentage during social activities, and 'alone' reports have the highest percentage during studying.}
    \label{fig:alone_vs_not_alone_contexts}%
\end{figure*}

In addition to the country-specific impact on the frequency and nature of social contexts among participants, a temporal effect was also evident. Figure~\ref{fig:social_context_over_day} illustrates how the distribution of social context reports varies by the hour of the day. \lakmal{Participants reported} to be predominantly alone or with their romantic partners during the morning hours. As the day progressed to lunch and dinner times, \lakmal{participants reported to have spent} their time (particularly eating) with their families. 

Figure~\ref{fig:alone_vs_not_alone_contexts} depicts the co-occurrence of specific moods (positive and negative valence), activities, and semantic locations with the social context (alone or not). Regarding concurrent activities and whether someone is alone or not, activities such as studying, listening to lectures, sleeping, and personal care were predominantly carried out alone. In contrast, activities involving social interaction, such as family care, eating, and nurturing one's social life, were typically done with others. \lakmal{Furthermore, regarding semantic location, participants reported having been alone at home on 125,270 instances, which could be influenced by the COVID-19 pandemic and the prevalence of remote work and study environments \cite{fried2022mental}. However, given the remote work culture that is prevalent even after the pandemic, we could expect these patterns to hold to an extent \cite{lund2021future, broom2021home, mass2021work}. In terms of gender-specific differences, there were no notable disparities between men and women. Men reported being alone in 54.6\% of all self-reports, while women reported to be slightly less alone, accounting for 49.1\% of all self-reports. Additionally, women reported to have been with their families, averaging 31.6\%, which is approximately 5\% more than men. All other distinctions between women and men were minor in the broader context. These findings align with existing research indicating that young men tend to experience more loneliness \cite{barreto2021loneliness} and tend to be alone more in certain contexts \cite{fry2021rising}. Prior work also discussed similar findings that women tend to seek more social ties compared to men \cite{tamminen2019living}. However, it is important to note that being alone could be associated with a wide range of factors, including culture, geography, age, etc., in addition to gender \cite{barreto2021loneliness}}.

\begin{table*}
  \caption{Top ten sensing features to discern alone or not alone social contexts, depending on the ascending value of p-value after Bonferroni correction, from mixed effects models.}
  \Description{This table lists the top ten sensing features for discerning 'alone' or 'not alone' social contexts, based on p-values after Bonferroni correction from mixed effects models, for individual countries and the dataset as a whole. It shows distinct patterns of app usage and activity types in different countries, such as negative associations of app usage with being alone in Denmark, contrasting with the patterns in Paraguay and Italy.}
  \label{tab:mixed_effects_model}
  \resizebox{\textwidth}{!}{
  \begin{tabular}{l lllll| l lllll}

    \cellcolor[HTML]{EDEDED} &
\cellcolor[HTML]{EDEDED} \rotatebox{60}{\textbf{feature}} &
\cellcolor[HTML]{EDEDED} \rotatebox{60}{\textbf{coefficient}} &
\cellcolor[HTML]{EDEDED} \rotatebox{60}{\textbf{std. error}}  &
\cellcolor[HTML]{EDEDED} \rotatebox{60}{\textbf{z-score}} &
\cellcolor[HTML]{EDEDED} \rotatebox{60}{\textbf{p-value}} &
\cellcolor[HTML]{EDEDED} & 
\cellcolor[HTML]{EDEDED} \rotatebox{60}{\textbf{feature}} &
\cellcolor[HTML]{EDEDED} \rotatebox{60}{\textbf{coefficient}} &
\cellcolor[HTML]{EDEDED} \rotatebox{60}{\textbf{std. error}}  &
\cellcolor[HTML]{EDEDED} \rotatebox{60}{\textbf{z-score}} &
\cellcolor[HTML]{EDEDED} \rotatebox{60}{\textbf{p-value}}
\\

\multirow{10}{*}{\rotatebox[origin=c]{90}{\textbf{Denmark}}} & 
app\_news\_and\_magazines &
-0.105 &
0.041 &
-2.537 &
0.011 &
\multirow{10}{*}{\rotatebox[origin=c]{90}{\textbf{Paraguay}}} & 
app\_finance &
-0.043 &
0.012 &
-3.468 &
0.000
\\

&
app\_casual &
-0.041 &
0.016 &
-2.492 &
0.012 &
&
notifications\_w/o\_duplicates &
-0.018 &
0.005 &
-3.248 &
0.001
\\

&
app\_simulation &
-0.180 &
0.074 &
-2.416 &
0.015 &
&
app\_productivity &
-0.000 &
0.000 &
-3.195 &
0.001
\\

&
app\_action &
-0.078 &
0.032 &
-0.241 &
0.015 &
&
location\_altitude &
-0.000 &
0.000 &
-3.175 &
0.001
\\

&
app\_books\_and\_reference &
-2.651 &
1.099 &
-2.412 &
0.015 &
&
activity\_in\_vehicle &
-0.004 &
0.001 &
-2.807 &
0.004
\\

&
app\_sports &
-0.582 &
0.241 &
-2.412 &
0.015 &
&
activity\_running &
-0.025 &
0.010 &
-2.588 &
0.009
\\

&
app\_adventure &
-0.109 &
0.045 &
-2.409 &
0.015 &
&
notifications\_removed &
-0.003 &
0.001 &
-2.577 &
0.009
\\

&
proximity\_std &
-0.027 &
0.011 &
-2.375 &
0.017 &
&
wifi\_number\_of\_devices &
-0.007 &
0.003 &
-2.466 &
0.013
\\

&
app\_business &
-0.017 &
0.007 &
-2.364 &
0.018 &
&
activity\_on\_bicycle &
-0.016 &
0.006 &
-2.405 &
0.016
\\

&
app\_maps\_and\_navigation &
-0.352 &
0.149 &
-2.354 &
0.018 &
&
wifi\_minimum\_rssi &
-0.006 &
0.002 &
-2.172 &
0.029
\\

\arrayrulecolor{Gray}
\midrule
       
\multirow{10}{*}{\rotatebox[origin=c]{90}{\textbf{UK}}} & 

user\_presense\_time &
-0.000 &
0.000 &
1.987 &
0.046 &
\multirow{10}{*}{\rotatebox[origin=c]{90}{\textbf{Mongolia}}} & 
app\_video\_players\_editors &
0.000 &
0.000 &
2.613 &
0.008
\\

&
proximity\_minimum &
-0.008 &
0.004 &
-1.975 &
0.048 &
&
app\_shopping &
-0.056 &
0.021 &
-2.609 &
0.009
\\

&
proximity\_mean  &
-0.018 &
0.009 &
-1.909 &
0.056 &
&
app\_travel\_and\_local &
-0.056 &
0.022 &
-2.500 &
0.012
\\

&
screen\_number\_of\_episodes &
0.011 &
0.006 &
-1.844 &
0.065 &
&
app\_trivia &
-0.674 &
0.284 &
-2.373 &
0.017
\\

&
app\_photography &
0.006 &
0.003 &
1.802 &
0.701 &
&
app\_books\_and\_reference &
-0.016 &
0.007 &
-2.312 &
0.020
\\

&
steps\_counter &
0.000 &
0.000 &
1.775 &
0.075 &
&
app\_educational &
-0.109 &
0.047 &
-2.300 &
0.021
\\

&
proximity\_maximum &
-0.015 &
0.009 &
-1.608 &
0.107 &
&
app\_weather &
-3.481 &
1.516 &
-2.295 &
0.021
\\

&
app\_entertainment &
0.020 &
0.012 &
1.588 &
0.112 &
&
app\_comics &
-0.379 &
0.165 &
-2.291 &
0.021
\\

&
activity\_on\_bicycle &
0.007 &
0.004 &
1.559 &
0.118 &
&
app\_art\_and\_design &
-0.705 &
0.315 &
-2.238 &
0.025
\\

&
app\_business &
0.003 &
0.002 &
1.440 &
0.149 &
&
app\_news\_and\_magazines &
-0.134 &
0.060 &
-2.227 &
0.025
\\
 
 \arrayrulecolor{Gray}
\midrule

\multirow{10}{*}{\rotatebox[origin=c]{90}{\textbf{Italy}}} & 
activity\_in\_vehicle &
0.004 &
0.000 &
5.970 &
0.000 &
\multirow{10}{*}{\rotatebox[origin=c]{90}{\textbf{All}}} & 
app\_news\_and\_magazines &
-0.105 &
0.041 &
-2.537 &
0.011 
\\

&
activity\_onfoot &
0.003 &
0.000 &
4.092 &
0.000 &
&
app\_casual &
-0.041 &
0.016 &
-2.492 &
0.012
\\

&
activity\_walking &
0.003 &
0.000 &
4.010 &
0.000 &
&
app\_simulation &
-0.180 &
0.074 &
-2.416 &
0.015 
\\

&
steps\_counter &
0.000 &
0.000 &
3.243 &
0.001 &
&
app\_action &
-0.078 &
0.032 &
-2.414 &
0.015
\\

&
activity\_tilting &
0.000 &
0.000 &
3.225 &
0.001 &
&
app\_books\_and\_reference &
-2.651 &
1.099 &
-2.412 &
0.015
\\

&
location\_altitude &
0.000 &
0.000 &
3.164 &
0.001 &
&
app\_sports &
-0.582 &
0.241 &
-2.412 &
0.015
\\

&
proximity\_mean &
-0.013 &
0.004 &
-3.067 &
0.002 &
&
app\_adeventure &
-0.109 &
0.045 &
-2.409 &
0.015
\\

&
notifications\_w/o\_duplicates &
-0.006 &
0.002 &
-2.409 &
0.015 &
&
proximity\_std &
-0.027 &
0.011 &
-2.375 &
0.017
\\

&
wifi\_std\_rssi &
0.004 &
0.001 &
2.394 &
0.016 &
&
app\_business &
-0.017 &
0.007 &
-2.364 &
0.018
\\

&
screen\_number\_of\_episodes &
0.003 &
0.001 &
2.111 &
0.034 &
&
app\_maps\_and\_navigation &
-0.352 &
0.149 &
-2.354 &
0.018
\\
 
\arrayrulecolor{Gray}
 \bottomrule
 
  \end{tabular}
  }
\end{table*}

\aurel{The statistical analysis presented in Table~\ref{tab:mixed_effects_model} highlights the top ten sensing features effective in discerning whether a person is alone or not, based on the p-value after Bonferroni correction. This analysis accounts for both fixed (features) and random effects (study participants), offering a nuanced understanding of the data. In Denmark, app usage features such as news and magazines, simulation, and books and references show moderate to high negative coefficients, indicating that using these apps is inversely related to being with others. Conversely, in Paraguay, app usage features like finance and productivity, along with the number of notifications received, have smaller negative coefficients. In the UK, features related to user presence and proximity are statistically significant but show almost no relationship with social context. Similarly, in Mongolia, features like trivia, comics, weather, and arts and design have notable negative relationships with being alone. Italy displays a unique pattern where physical activities such as being in a vehicle, on foot, and walking, though statistically significant, have very small coefficients, showing no clear association. This pattern, at least in terms of feature set, is akin to the UK and Paraguay but contrasts with Denmark and Mongolia, where app usage features have the top statistical significance values. It is also worth noting that, in the UK, except for two features (user\_presense\_time and proximity\_minimum), other features had $p \geq 0.05$, suggesting no statistical significance. The `All' category, aggregating data from all countries, indicates that app features like news and magazines, books and references, maps, sports, and simulation are generally negatively associated with being alone. These findings highlight country-specific differences in the relationship between sensor data features and social context, emphasizing the need to consider user behavior diversity in different countries for understanding and inferring everyday life behaviors with multimodal sensor data \cite{khwaja2019modeling, assimeegahapola2023complex}.}

\subsection{\textbf{RQ2: Social Context Inference Without Personalization}\label{section:social_cont_infer}}

This section addresses the question of whether social context can be inferred using smartphone sensing data (\textbf{RQ2}). A comparison of the results obtained is presented in Table~\ref{tab:model_selection}. Overall, in the generic multi-country setup, random forest models demonstrated the best performance. This performance was approximately 2\% higher than that of XGBoost. Consequently, all subsequent results are only shown using random forest models for the sake of brevity. The test \aurel{AUC of the random forest models ranged from 63\% to 73\%,} depending on the country. However, due to the relatively small dataset sizes in some countries (e.g., only 10K self-reports in Denmark and 9.7K self-reports in Paraguay) and the high number of features, we employed feature selection to explore whether performance could be improved.

\begin{table*}[t] 
  \caption{Different machine learning models and their inference AUC and F1 metrics with standard deviations in brackets according to different countries.}
  \Description{The table displays the performance of various machine learning models, including their AUC and F1 scores with standard deviations, in predicting social context across different countries. It highlights that random forest classifiers generally achieved the best performance in terms of both AUC and F1 scores across many countries.}

  \label{tab:model_selection}
  \resizebox{0.95\textwidth}{!}{
  \begin{tabular}{lllllllll}
    \cellcolor[HTML]{EDEDED} &
    \multicolumn{2}{l}{\cellcolor[HTML]{EDEDED} \textbf{Logistic (L2)} } &
    \multicolumn{2}{l}{\cellcolor[HTML]{EDEDED} \textbf{Random Forest} } &
    \multicolumn{2}{l}{\cellcolor[HTML]{EDEDED} \textbf{XG Boost} } &
    \multicolumn{2}{l}{\cellcolor[HTML]{EDEDED} \textbf{Ada Boost} }
    \\
    
    \cellcolor[HTML]{EDEDED} &
    \cellcolor[HTML]{EDEDED} \textbf{$\overline{AUC}$ ($AUC_\sigma$)}  & 
    \cellcolor[HTML]{EDEDED} \textbf{$\overline{F1}$ ($F1_\sigma$)} &
    \cellcolor[HTML]{EDEDED} \textbf{$\overline{AUC}$ ($AUC_\sigma$)} & 
    \cellcolor[HTML]{EDEDED} \textbf{$\overline{F1}$ ($F1_\sigma$)} &
    \cellcolor[HTML]{EDEDED} \textbf{$\overline{AUC}$ ($AUC_\sigma$)} &
    \cellcolor[HTML]{EDEDED} \textbf{$\overline{F1}$ ($F1_\sigma$)} &
    \cellcolor[HTML]{EDEDED} \textbf{$\overline{AUC}$ ($AUC_\sigma$)} &
    \cellcolor[HTML]{EDEDED} \textbf{$\overline{F1}$ ($F1_\sigma$)} \\


Majority Baseline & 50.0 (0.0)  & 33.7 (0.1) & 50.0 (0.0) & 33.7 (0.1) & 50.0 (0.0)  & 33.7 (0.1) & 50.0 (0.0) & 33.7 (0.1)  \\
Random Baseline & 50.1 (0.1)  & 50.0 (0.2) & 50.1 (0.1) & 50.0 (0.2) & 50.1 (0.1)  & 50.0 (0.2) & 50.1 (0.1) & 50.0 (0.2)  \\

Multi-Country & 50.5 (2.8) &  36.6 (9.5) & \textbf{67.6} (4.1) &   \textbf{61.3} (2.8) & 65.8 (2.3)  &  60.5 (1.7) & 61.0 (2.8) &  56.1 (3.2) \\
\arrayrulecolor{Gray}
\midrule 

UK & 49.5 (2.7) &  41.6 (7.1) & \textbf{63.5 }(4.5) &  \textbf{55.7} (4.6) & 60.8 (5.2) &  55.0 (3.3) & 59.3 (6.5) &  54.9 (5.1) \\
Denmark & 51.1 (1.0) &  32.7 (10.0) & \textbf{66.0} (8.7) &  55.0 (9.8) & 60.3 (9.5) &  53.3 (8.8) & 61.2 (8.8)  &  \textbf{56.1} (7.5) \\
Italy & 49.9 (0.5) &  29.3 (2.4) & \textbf{66.6} (1.9) &  \textbf{60.9 }(1.2) & 66.1 (2.6)  &  60.8 (2.3) & 63.8 (2.8) &  59.2 (2.0) \\ 
 Paraguay & 49.7 (1.4) &  37.1 (5.9) & \textbf{72.0 }(12.8) &  63.1 (9.4) & 70.9 (11.8) & \textbf{64.1} (9.3) & 65.7 (13.4) &  60.7 (10.5) \\
 Mongolia & 49.5 (2.0) &  43.5 (1.8) & \textbf{73.8 }(6.2) &  65.3 (6.1) & 71.9 (5.6) &  \textbf{66.4 }(4.9) & 63.9 (5.3) &  58.4 (3.9) \\
\bottomrule
\end{tabular}
}
\end{table*}

\aptLtoX[graphic=no,type=html]{\begin{figure}[t]
        \centering
        \includegraphics[width=\linewidth]{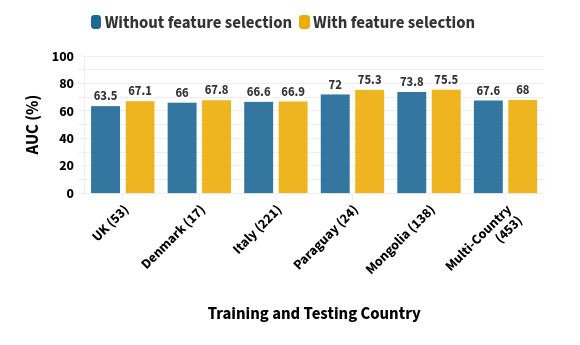}
        \caption{Comparison of random forest model performance for feature selection and without feature selection.\aurel{ The performance metric shown is AUC.}}
        \Description{This figure compares the performance, in terms of AUC, of random forest models with and without feature selection. It shows that multi-country models perform similarly to country-specific models.}
        \label{fig:feature_selection_auc}
    \end{figure}
    \begin{figure}
        \centering
        \includegraphics[width=\linewidth]{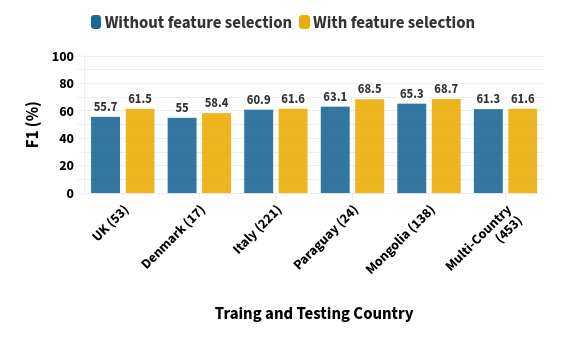}
        \caption{Comparison of random forest model performance for feature selection and without feature selection. The performance metric shown is F1 score.}
        \Description{This figure similarly compares the performance of random forest models with and without feature selection, but in terms of the F1 score. It shows that multi-country models are on par with country-specific models.}
        \label{fig:feature_selection_f1}
\end{figure}
\begin{figure}
        \centering
        \includegraphics[width=\linewidth]{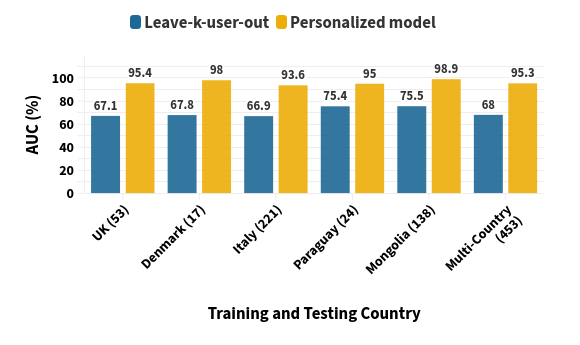}
        \caption{Comparison of random forest model performance for population-level (leave-k-out) and hybrid (personalized) models. The performance metric shown is AUC.}
        \Description{This figure compares the performance, measured by AUC, of random forest models for population-level (leave-k-out) and hybrid (personalized) models. It demonstrates that hybrid models achieve high AUCs above 90\% in both country-specific and multi-country setups.}
        \label{fig:model_personalization_auc}
    \end{figure}
    \begin{figure}
        \includegraphics[width=\linewidth]{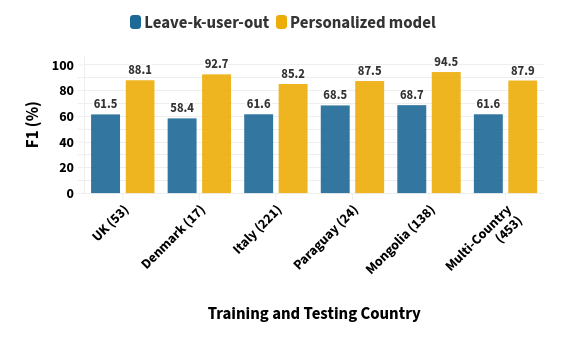}
        \caption{Comparison of random forest model performance for population-level (leave-k-out) and hybrid (personalized) models. The performance metric shown is the F1 score.}
        \Description{This figure compares the performance, measured by F1 score, of random forest models for population-level (leave-k-out) and hybrid (personalized) models. It shows that hybrid models achieve high F1 scores above 85\% in both country-specific and multi-country setups.}
        \label{fig:model_personalization_f1}
    \end{figure}}{\begin{figure*}[t]
\begin{center}
    \begin{minipage}[t]{0.45\textwidth}
        \centering
        \includegraphics[width=\linewidth]{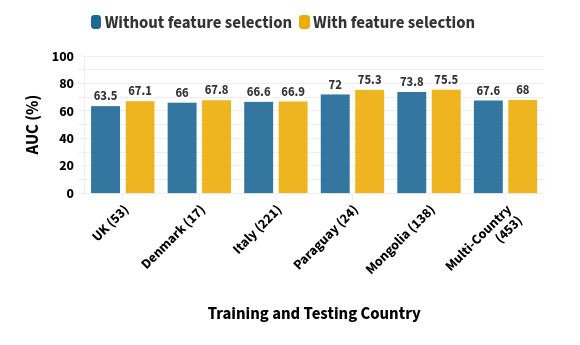}
        \caption{Comparison of random forest model performance for feature selection and without feature selection.\aurel{ The performance metric shown is AUC.}}
        \Description{This figure compares the performance, in terms of AUC, of random forest models with and without feature selection. It shows that multi-country models perform similarly to country-specific models.}
        \label{fig:feature_selection_auc}
    \end{minipage}
    \hfill
    \begin{minipage}[t]{0.45\textwidth}
        \centering
        \includegraphics[width=\linewidth]{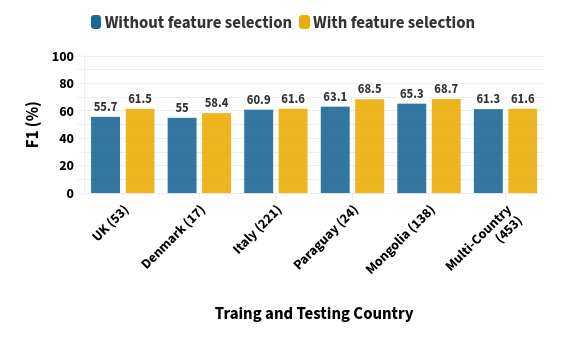}
        \caption{Comparison of random forest model performance for feature selection and without feature selection. The performance metric shown is F1 score.}
        \Description{This figure similarly compares the performance of random forest models with and without feature selection, but in terms of the F1 score. It shows that multi-country models are on par with country-specific models.}
        \label{fig:feature_selection_f1}
    \end{minipage}
\end{center}


\begin{center}
    \begin{minipage}[t]{0.45\textwidth}
        \centering
        \includegraphics[width=\linewidth]{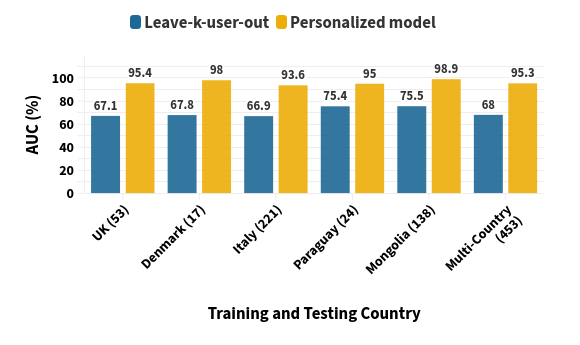}
        \caption{Comparison of random forest model performance for population-level (leave-k-out) and hybrid (personalized) models. The performance metric shown is AUC.}
        \Description{This figure compares the performance, measured by AUC, of random forest models for population-level (leave-k-out) and hybrid (personalized) models. It demonstrates that hybrid models achieve high AUCs above 90\% in both country-specific and multi-country setups.}
        \label{fig:model_personalization_auc}
    \end{minipage}
    \hfill
    \begin{minipage}[t]{0.45\textwidth}
        \centering
        \includegraphics[width=\linewidth]{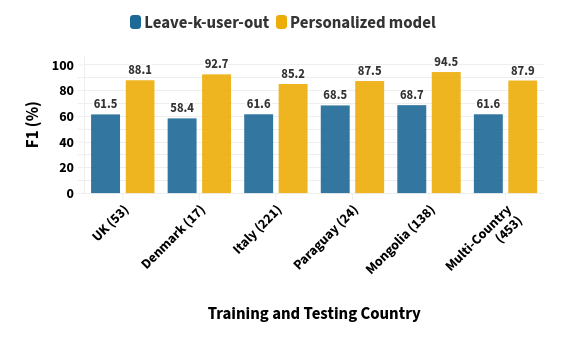}
        \caption{Comparison of random forest model performance for population-level (leave-k-out) and hybrid (personalized) models. The performance metric shown is the F1 score.}
        \Description{This figure compares the performance, measured by F1 score, of random forest models for population-level (leave-k-out) and hybrid (personalized) models. It shows that hybrid models achieve high F1 scores above 85\% in both country-specific and multi-country setups.}
        \label{fig:model_personalization_f1}
    \end{minipage}
\end{center}
\vspace{-0.2 in}
\end{figure*}}

%

\aurel{In Figure~\ref{fig:feature_selection_auc} and Figure~\ref{fig:feature_selection_f1}, we compare the AUC and F1-Score, respectively,} of different country-specific models with and without feature selection. The improvements in \aurel{ AUC vary depending on the country and, consequently, the number of participants, ranging from 0.3\% (Italy) to 3.5\% (UK). Considering the F1 metric slightly bigger improvements can be observed, which range from around 1\% (Italy) up to 5.5\% (UK). } This relatively modest performance increase can be attributed to the fact that random forest classifiers already incorporate embedded feature selection during the training phase, making explicit feature selection prior to training less impactful \cite{meegahapola2023generalization}. Even when examining other model types, we did not observe notable improvements in feature selection. Additionally, the optimal number of features varied by country but tended to hover slightly above 30. Specifically, the final feature counts used in training were 31 in the UK, 32 in Denmark, 36 in Italy, 30 in Paraguay, 36 in Mongolia, and 90 in the Multi-Country approach. Hence, compared to country-specific models with lower feature counts, feature selection in the multi-country approach retained over 90 features, suggesting that when data from diverse countries are used to train a single model, more features are needed for better representation. \lakmal{It is also important to note that this outcome might not just be due to the number of data points or users in the training data. Countries with a higher number of users (Italy, Mongolia) still resulted in models with fewer features after feature selection, similar to countries with a lower number of data points (UK, Denmark, Paraguay). Therefore, this result is logical and should be expected when using larger behavioral datasets from diverse countries.} This underscores the notion that different cultural and social practices in various countries impact the performance of social context inference models, as revealed in the analysis presented in Table \ref{tab:mixed_effects_model}.


\subsection{\textbf{RQ3: Effect of Partial Personalization on Model Performance}}

Next, our exploration delves into the impact of model personalization on performance. In Figure~\ref{fig:model_personalization_auc} and Figure~\ref{fig:model_personalization_f1}, we draw comparisons between population-level and hybrid models across individual countries and the multi-country approach. These results are derived after feature selection. First and foremost, our findings underscore the importance of personalization, as it consistently shows improved performance across all countries and the multi-country approach, with performance gains \aurel{of the range 20-30\% in AUC}. This aligns with prior research in domains like mood inference and eating behavior prediction, where smartphone sensing-based models have demonstrated enhanced performance following personalization \cite{likamwa2013moodscope, meegahapola2023generalization, bangamuarachchi2022sensing, bouton2022your}. Thus, it comes as no surprise that we observe a similar trend in the context of social context inference. Hence, the notable performance boost in personalized models highlights the individualistic and variable nature of social context, prompting further exploration into conceptualizing and modeling behavioral diversity within sensing models.

In fact, it is worth noting that the weighted average AUC (weighted by the number of data points) of the country-specific models slightly outperforms the population-level multi-country model, achieving \aurel{68.7\% compared to 67.6\%, albeit by a narrow margin. This trend is consistent across hybrid models as well.} 
Nevertheless, it is important to consider that the standard deviations, calculated based on ten iterations of experiments as explained in Section \ref{sec:experimentalsetup}, reveal a larger variability in the performance of country-specific models \aurel{}(7.1\%) compared to the multi-country model (4.0\%). This discrepancy can be partly attributed to the relatively small number of participants in certain countries, such as Denmark and Paraguay. Furthermore, the hybrid models exhibit good performance, surpassing \aurel{90\% AUC across all individual countries and in the multi-country setup}. This outcome underscores the effectiveness of model personalization in enhancing the performance of social context inference models. In conclusion, our findings indicate that training a multi-country model yields reasonably good results for social context inference. This result diverges from prior research that explored the performance of country-specific and multi-country models across mood inference, personality inference, and activity recognition \cite{khwaja2019modeling, meegahapola2023generalization, assimeegahapola2023complex}, where multi-country models tended to under-perform by larger margins. Hence, it is important to highlight that this distinction does not extend to social context inference, where the multi-country approach proves to be effective. \lakmal{On a broader scheme of things, these results also suggest that there is feasibility in building models for social context inference that capture the complex everyday life behavior of individuals well.}

\lakmal{The multi-country approach, with its access to a larger data volume, includes diverse country-specific characteristics. This diversity makes it challenging to definitively attribute the differences in inference performance between multi-country and country-specific models to either data quantity or country-specific data heterogeneity. Prior research in personality modeling and mood inference has grappled with the complexity of providing a clear answer in such scenarios involving multi-country data. Employing under-sampled models, which use an equal amount of data from each country to train country-specific models (often matched to the data volume of the country with the least data), could inadvertently result in reduced representational expressiveness for individual countries \cite{khwaja2019modeling, meegahapola2023generalization}.}

\subsection{\textbf{RQ4: Generalization of Models to Unseen Countries}}

\begin{table*}[t] 
  \caption{Country-agnostic: Using the countries in the left column, to predict data from the countries on the top row, with AUC, F1 and their tenfold standard deviation. Values along the diagonal are grayed out as they came from the performance obtained in Figure \ref{fig:feature_selection_auc}, with feature selection under the country-specific approach.}
  \Description{This table shows the performance of country-agnostic models in predicting social context in different countries, using AUC and F1 scores with their standard deviations. The diagonal values, which are grayed out, indicate the performance obtained from the country-specific approach with feature selection, as shown in Figure 6. Results show that social context inference models do not generalize well to other countries.}
  \label{tab:country_agnostic_matrix}
  \resizebox{0.95\textwidth}{!}{
  \begin{tabular}{llllll}
    \cellcolor[HTML]{EDEDED} \textit{Testing} &
    \cellcolor[HTML]{EDEDED} \textbf{UK}  &
    \cellcolor[HTML]{EDEDED} \textbf{Denmark} &
    \cellcolor[HTML]{EDEDED} \textbf{Italy} &
    \cellcolor[HTML]{EDEDED} \textbf{Paraguay} &
    \cellcolor[HTML]{EDEDED} \textbf{Mongolia} \\

    \cellcolor[HTML]{EDEDED} \textit{Training} &
     &
    &
    &
    &
    \\

\cellcolor[HTML]{EDEDED} \textbf{UK}  & \textcolor{gray}{67.1 (4.0) / 61.5 (3.2) } & 	54.2 (10.0) / 50.1 (8.1)  & 60.3 (3.1) / 56.7 (1.8)  & 61.8 (8.9) / 55.6 (6.9)  & 53.8 (3.9) / 42.5 (5.9) \\
\cellcolor[HTML]{EDEDED} \textbf{Denmark} & 51.9 (4.2) / 47.0 (3.8)  & \textcolor{gray}{67.8 (9.2) / 58.4 (9.6)} & 	55.4 (3.3) / 49.0 (2.7)  & 56.0 (7.8) / 51.7 (5.8)  & 52.2 (6.8) / 50.7 (3.9) \\
\cellcolor[HTML]{EDEDED} \textbf{Italy} & 61.9 (4.3) / 55.8 (1.8)  & 60.4 (10.8) / 54.5 (11.8) & 	\textcolor{gray}{66.9 (2.6) / 61.5 (2.0)} & 	54.2 (6.6) / 50.8 (7.0)  & 57.6 (5.1) / 42.0 (8.5) \\
\cellcolor[HTML]{EDEDED} \textbf{Paraguay} &  58.0 (7.1) / 51.8 (7.5)  & 50.1 (11.2) / 43.8 (9.4)  & 56.1 (3.3) / 53.1 (2.1) & \textcolor{gray}{75.4 (14.4) / 68.5 (13.0) } & 	57.4 (5.1) / 48.8 (4.5) \\
\cellcolor[HTML]{EDEDED} \textbf{Mongolia} &55.5 (6.8) / 31.0 (11.1)  & 54.4 (13.9) / 45.4 (8.4)  & 52.8 (3.7) / 34.8 (4.5)  & 62.8 (9.8) / 47.5 (13.7) & 	\textcolor{gray}{75.5 (5.4) / 68.7 (6.1)} \\

\arrayrulecolor{Gray}
\bottomrule

\end{tabular}
}
\end{table*}

\begin{figure}[t]
    \centering
    \includegraphics[width=\linewidth]{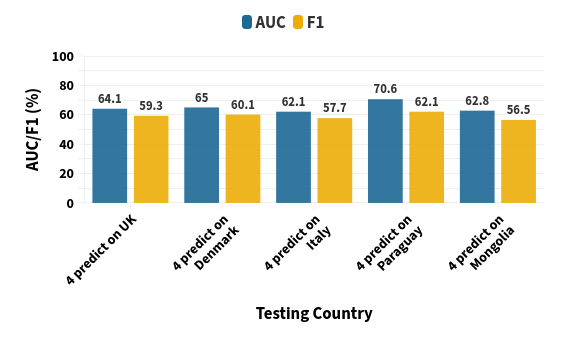}
    \caption{Country-agnostic model performance with random forest classifiers, for different testing countries where models were trained on the other four countries. AUC and F1 score is shown.}
    \Description{This figure illustrates the performance of country-agnostic models using random forest classifiers, with AUC and F1 scores for different testing countries. It highlights that these models consistently score lower in AUC compared to country-specific models, though the differences are generally minor.}
    \label{fig:country_agnostic}
\end{figure}

To gain a deeper understanding of how data heterogeneity tied to a particular country aids in detecting the behavior of its participants and the impact of additional data on \aurel{inference performance}, we conducted another experiment. Here, we utilized data from four countries to test it in a fifth, unseen country. This setup emulates a scenario where an application is developed and trained for a specific set of countries and then enters a new market in a different country. The results of this experiment are illustrated in Figure \ref{fig:country_agnostic}. Notably, the AUC of these country-agnostic models consistently falls below those of the country-specific models, although the differences are minor. \aurel{The magnitude of the AUC variance between country-agnostic and country-specific approaches is minimal in Denmark and the UK ($\Delta$ 3\%), the country with the least available data, and most pronounced in Mongolia ($\Delta$ 12.7\%), a country with high data volume.}

Lastly, we explored another experimental setting in which data from one country were used to train a model, which was then tested separately in all the other countries. The results are presented in Table \ref{tab:country_agnostic_matrix}. This setup allowed us to investigate the geographical proximity between different countries and how it translates to social context inference models. In the resulting matrix, the rows represent the training countries, while the columns represent the testing country (e.g., using UK data to test in Denmark yields a \aurel{54.2\% AUC}). Interestingly, culturally similar countries appeared to extrapolate better to each other in certain cases. For instance, it can be observed that Italian model worked comparatively well for the UK and Denmark, while it did not perform as well for Paraguay or Mongolia. However, this pattern did not hold to models trained in Denmark and the UK when tested in other European countries. This finding contrasts with prior research on mood inference and complex daily activity recognition in a multi-country setting \cite{meegahapola2023generalization, assimeegahapola2023complex}, where the authors found models performing reasonably well when tested in geographically similar European countries. However, similar to those studies and another study \cite{Xu2023Globem}, we observed a performance degradation, which could be attributed to distributional shifts between countries. Furthermore, Mongolia appeared to be a challenging country to predict without Mongolian data in the training set. \aurel{If we tested a model in Mongolia using Mongolian data, we could achieve a mean AUC of 75.5\%. However, using training data from any other country, the highest AUC achieved was 57.6\%, with a model trained on Italian data, marking a difference of approximately 18\%. For other countries, the differences ranged between 5\% and 15\%.} This disparity could be attributed to either distributional differences in sensor data across countries or variations in the target attribute, i.e., the social context. In fact, as seen in Figure~\ref{fig:social_context_countries}, the distribution of the social context in Mongolia differed from that of all other countries, featuring fewer alone self-reports. \lakmal{Additionally, as shown in Table~\ref{tab:mixed_effects_model}, features with the highest statistical significance for Mongolia were predominantly app usage-based, a pattern that was distinct from other countries except Denmark.} Therefore, this could potentially explain the disparity in inference performance in this context.

\section{Discussion}\label{sec:discussion}

\subsection{Summary of Results}\label{subsec:summary}
In summary, our results shed light on the research questions as follows:

\begin{itemize}[wide, labelwidth=!, labelindent=0pt]
    \item \textbf{RQ1}: We conducted a comprehensive analysis of a smartphone sensing dataset collected from 483 young adults across five countries. Our analysis identified specific individual features, such as types and quantities of app usage (including tools, communication, productivity, and social), activity types (like being in a vehicle, on foot, walking), location, Bluetooth and WiFi usage, proximity, and screen episodes, as top indicators for distinguishing between being alone and not alone. Moreover, we found that the features significant in differentiating social contexts varied across countries, emphasizing the diversity in behaviors and contextual features associated with various social settings.

    \item \textbf{RQ2}: We operationalized three approaches to assessing the inference of social context (as described in Table~\ref{tab:terminology})—these approaches were country-specific, country-agnostic, and multi-country. Without accounting for the diversity of data sources at the country level, we found that the generic multi-country approach exhibits a moderate AUC of 67.6\% without feature selection and 68\% with feature selection when using population-level models. The country-specific approach resulted in \aurel{similar} performance to the multi-country approach, with most countries displaying margins of less than 5\% for both with and without feature selection. This outcome does not follow prior studies that utilized similar multi-country datasets for health-related inferences, which suggested that country-specific models were generally superior by substantial margins. As a result, we concluded that this may not always hold true for social context inference.

    \item \textbf{RQ3}: The utilization of hybrid models resulted in the multi-country approach yielding an AUC of 95.3\%. Country-specific models performed similarly, with their differences in AUC being in the range of 3\%. \aurel{The results indicate that a generic multi-country model may generally suffice for social context inference, regardless of personalization level. Country-specific models might offer slight improvements in certain countries (i.e., UK, Denmark, and Mongolia). Notably, this deviates from previous research that use smartphone sensor data for health-related inferences, where country-specific models were recommended to achieve better performance, even when personalized. This contrast highlights the differing effectiveness of model types across various domains and objectives.}

    \item \textbf{RQ4}: With regards to mood inference and activity recognition, despite initial prior research \cite{meegahapola2023generalization, assimeegahapola2023complex} indicating that models may exhibit reasonable generalization to geographically proximate countries in Europe, our analysis of social context inference performance did not reveal such associations. In fact, in all cases, the models performed poorly when applied to unseen countries. This finding is consistent with prior studies, which concluded that models lack generalization capabilities when deployed to previously unseen countries compared to the performance of country-specific or multi-country approaches. \lakmal{In practice, this means that social context inference models could either be developed in deployment countries or developed with data collected from multiple countries by capturing a broader behavioral diversity, leading to robust models.}

\end{itemize}

\subsection{Implications of Social Context Inference for Future Applications}\label{subsec:implications}

This paper discusses the operationalization and achieved performance of a social context inference task in a multi-country setting. The presented work offers theoretical and practical implications for researchers and practitioners in human-computer interaction and ubiquitous computing.

The theoretical implications of the present study are manifold. Firstly, we propose an experimental framework for implementing and comparing social context inference experiments in a multi-country context. The conducted experiments include country-specific, country-agnostic, and multi-country approaches, offering insights into the extent to which country-specific social context information is encoded in multimodal sensing data and whether this information can enhance model performance. In future human sensing research exploring multi-country settings, this experimental framework may be adopted to ensure comparability across studies, facilitating cross-dataset comparisons. Secondly, our findings indicate that country-specific models outperform multi-country models by only a marginal degree. This does not follow previous research on mood inference and activity recognition, which suggested that country-specific models are superior to multi-country models. However, given the complexity of the data collected during the COVID-19 pandemic and other data limitations, the results presented here are not definitive. Consequently, future studies should examine the impact of country diversity on model performance in context-aware sensing tasks, paying attention to cultural practices that may drive differences in social behavior and that may be apparent in the multimodal data. Thirdly, our results demonstrate the importance of personalization for the social context inference task. Hybrid models outperform population-level models by a large margin, suggesting that users exhibit significant differences in social behavior. Consequently, future studies should take note of the significance of collecting sufficient data from individual users to enable personalized models to be implemented.

\lakmal{The practical implications of our study in social context inference extend to various real-world applications. Firstly, mobile health applications that include social context models need to address the initial low inference performance which might lead to inappropriate notifications for new users. One strategy, similar to some mobile health apps that delay providing insights due to lack of personalization \cite{meegahapola2020protecting}, is to initially disable the social context feature and enable it only after the model has been adequately personalized for an individual user. Secondly, applications should be designed to monitor social context over an extended period, enabling behavior and health monitoring apps to provide users with long-term insights into their social patterns, such as the amount of time spent alone, empowering users to make informed decisions about their social behavior. This could have implications on their mental well-being. Thirdly, employing social context models that are pre-trained on large datasets from various countries and subsequently personalized with user-provided data can ensure high accuracy in inferring individual social contexts. As a fourth implication, these models have notable potential in smart environments and wearable technology. By understanding the social context, smart devices and environments can adapt functionalities to enhance user experience. For instance, a smart home system could modify its ambiance, such as lighting and music, based on whether the user is alone or hosting guests. Similarly, wearable devices and phones can adjust their notification settings---vibrating for notifications when the user is in a meeting and switching to visual or audible alerts when they are alone. This adaptation not only improves the user's interaction with technology but also respects social norms and privacy preferences. In educational and workplace settings, this technology can be used to tailor communication strategies and content delivery. For instance, learning platforms can adjust content presentation based on the user’s social environment, and workplace tools can optimize task allocation and collaboration features to align with the user's current social context. This wide range of implications underscores the extensive potential of social context inference across daily life situations in improving interactions, user experience, health, and well-being.}

\subsection{Cross-Cultural Considerations for Human-Computer Interaction Researchers}

Our study also has implications for HCI researchers, particularly in the realm of global applications. It underscores the critical importance of incorporating cross-cultural considerations into the design, development, and deployment of context-aware systems. Understanding that social behaviors and contextual features can significantly differ between countries and cultures emphasizes the need for adaptable and culturally sensitive models. HCI researchers should recognize the wide-ranging variations in individual behaviors, preferences, and social norms across diverse cultural contexts. Consequently, when developing context-aware applications, conducting extensive cultural research and adapting algorithms and models to local practices and expectations is essential. This not only acknowledges the diversity of behaviors but also respects the ethical norms inherent to each culture. Moreover, achieving a balance between personalization and universality is a central challenge. While personalization is critical for accurate social context inference, there is an increasing need for models and systems that can function across different cultural settings. Creating models that can adapt to varying cultural contexts without compromising user privacy or \aurel{model performance} remains a significant goal. In summary, recognizing the profound impact of cross-cultural diversity on context-aware systems is an imperative consideration for HCI researchers. By embracing this perspective, researchers can craft more culturally sensitive applications that enhance user experiences while respecting country-specific cultural norms.

\subsection{Ethical and Privacy Considerations of Social Context Inference}

This study sheds light on the important ethical and privacy concerns that arise with the implementation of context-aware systems, especially with social context inference. Researchers and practitioners are faced with a complex challenge: how to provide valuable services while also respecting user privacy and autonomy. These findings underscore the importance of informed consent and user control in the development of applications that perform social context inference. Transparency should be a top priority when creating such applications. Users must have a clear understanding of how their data is collected, processed, and used to infer social context. Additionally, they should be granted precise control over when and how this inference takes place. Empowering users to make informed decisions regarding context-aware features, especially those involving social context, is crucial, given that this information not only concerns the individual, but also those in their proximity. Furthermore, privacy-preserving techniques and data anonymization should be at the core of the design of context-aware systems. Even though the social context may not constitute health-related information, it holds great relevance to individuals and is closely linked to various aspects of health. Moreover, researchers should be attuned to the variations in privacy expectations and requirements in different countries, adhering to international privacy standards such as the General Data Protection Regulation (GDPR) and local regulations to ensure ethical and legal compliance. Another critical aspect is the potential for bias and discrimination within context-aware applications. Researchers must be vigilant in identifying and mitigating biases that could arise during social context inference. Regular audits and assessments should be conducted to detect and rectify any discriminatory outcomes associated with the inferred social behavior. Finally, researchers have a pivotal role in advocating for robust ethical guidelines and regulations governing the development and deployment of context-aware systems, including social context. Collaboration with policymakers, privacy advocates, and industry stakeholders is essential to establish ethical frameworks that safeguard user privacy and uphold ethical standards within the field. 

\section{Limitations and Future Work}\label{subsec:limitations}

We conclude by discussing the limitations of and future work resulting from our study. These concern the data generation process, experimental setup, and the computational approaches used. \lakmal{First, a primary limitation of our study is the use of the term "country" to refer to the student populations in universities within those countries, which may not represent the entire country. Although we have been transparent about this from the beginning of the paper, it is important to interpret the results cautiously, keeping this aspect in mind}. All data used for this study was generated in a time frame from November to December 2020. This coincides with a major surge of the COVID-19 pandemic in Europe but also in Paraguay and Mongolia. In all countries, social distancing measurements were recommended. Obviously, the COVID pandemic deeply influenced and possibly altered the behaviors, social practices, and moral norms of the pilot study participants. It is, however, highly difficult to assess how those changes influenced the analyzed data. It is quite likely that the pandemic changed the frequency of social contexts (fewer reports being done with others), but it is impossible to say if and how it precisely changed social interactions if they occurred and how this, on the other hand, influenced the mobile sensing data produced. On the other hand, given that many universities and companies are already adhering to remote work/study settings, we expect this trend to continue for the years to come. Hence, despite the study being done during the pandemic, we expect these results to hold in the future to an extent. However, future studies could explore this further.

This study considers only the two-class social context inference task of predicting if someone is alone or not. In principle, however, the used dataset differentiates between 8 different social contexts. The decision to limit this analysis to a two-class inference is mostly caused by class imbalance and page limits. First of all, the alone or not alone situation is well populated in all country data subsets, which is not necessarily given for any other two-class social contexts (e.g., with family, without family would not be well represented in Denmark). Thus, to obtain comparable results across all countries, only this most naive social context was exhaustively investigated. Furthermore, feasibility reasons also tempered further experiments with other social context inference tasks. The performance of the most simple social inference task lies at minimally \aurel{66.9\% AUC}, in the case of, for example, Italy, already quite low; thus, it can be assumed that models would perform even worse on more difficult social inference tasks. Future work could explore other constructs of the social context. \lakmal{Additionally, while we experimented with various time windows, we have only reported results using a ten-minute window for aggregating sensor data. This was done to facilitate easier comparison with other studies that used the same dataset \cite{assimeegahapola2023complex, meegahapola2023generalization}. However, it is important to note that the choice of time window depends on the nature of the dataset and the specific inference task. While the ten-minute window was suitable for our tasks, including social context inference, different time windows could be explored in future work.}

The results presented in the personalization section show that social behavior differs greatly from user to user. This indicates that even leave-k-users-out models could be greatly improved if user profiles or specific user groups could be more precisely modeled. This study tries to model different subgroups of users according to their country of residence and assumes similar cultural and social behaviors of people living in the same country. However, with additional information provided by users regarding other diversity attributes, models could potentially be further improved \cite{schelenz2021theory}. This would open the possibility of constructing partially personalized models on a set of diverse information provided by a new user. Future work should try to leverage the information on user diversity to improve mobile sensing models. \lakmal{Finally, it is important to note that although our study aimed to raise awareness about the diversity and generalization issues in modeling human behavior, our sample was primarily composed of young adults from various countries. While our research contributes positively to the broader conversation about generalization, we acknowledge that it also faces its own limitations, particularly regarding the lack of different age groups, occupations, and racial profiles.}

\section{Conclusion}\label{sec:conclusion}

\lakmal{In our study, we conducted an in-depth analysis to understand the social context of young adults across five countries using smartphone data. We compared the performance of models trained on data specific to each country against a model trained on data from all countries combined. Interestingly, we found that the country-specific models slightly outperformed the multi-country model, despite having less data. Additionally, our results showed that hybrid models were effective, indicating that a general model using data from various countries can successfully infer social context, with country-specific models offering minor performance improvements in certain cases. These findings have several practical applications. Mobile health apps incorporating social context models can provide valuable information, feedback, and interventions, allowing users to track their social context over time and gain insights into their well-being. In smart homes and wearable devices, social context inference models can adapt functionalities based on whether the user is alone or with others, enhancing the user experience. Our study highlights the potential of using smartphone data to passively infer social context, opening up various application opportunities. Furthermore, we discussed the importance of training these models with country diversity in mind to create robust systems that can adapt to diverse daily life situations across multiple countries.}

\begin{acks}
This work was funded by the European Union’s Horizon 2020 WeNet
project, under grant agreement 823783. We deeply thank all
the volunteers across the world for their participation in the study.
\end{acks}

\bibliographystyle{ACM-Reference-Format}
\bibliography{08_citations}


\begin{thebibliography}{97}


\ifx \showCODEN    \undefined \def \showCODEN     #1{\unskip}     \fi
\ifx \showDOI      \undefined \def \showDOI       #1{#1}\fi
\ifx \showISBNx    \undefined \def \showISBNx     #1{\unskip}     \fi
\ifx \showISBNxiii \undefined \def \showISBNxiii  #1{\unskip}     \fi
\ifx \showISSN     \undefined \def \showISSN      #1{\unskip}     \fi
\ifx \showLCCN     \undefined \def \showLCCN      #1{\unskip}     \fi
\ifx \shownote     \undefined \def \shownote      #1{#1}          \fi
\ifx \showarticletitle \undefined \def \showarticletitle #1{#1}   \fi
\ifx \showURL      \undefined \def \showURL       {\relax}        \fi
\providecommand\bibfield[2]{#2}
\providecommand\bibinfo[2]{#2}
\providecommand\natexlab[1]{#1}
\providecommand\showeprint[2][]{arXiv:#2}

\bibitem[Abdullah et~al\mbox{.}(2012)]%
        {abdullah2012towards}
\bibfield{author}{\bibinfo{person}{Saeed Abdullah}, \bibinfo{person}{Nicholas Lane}, {and} \bibinfo{person}{Tanzeem Choudhury}.} \bibinfo{year}{2012}\natexlab{}.
\newblock \showarticletitle{Towards population scale activity recognition: A framework for handling data diversity}.
\newblock \bibinfo{journal}{\emph{Proceedings of the AAAI Conference on Artificial Intelligence}} \bibinfo{volume}{26}, \bibinfo{number}{1}, \bibinfo{pages}{851--857}.
\newblock


\bibitem[Adams et~al\mbox{.}(2008)]%
        {adams2008sensing}
\bibfield{author}{\bibinfo{person}{Brett Adams}, \bibinfo{person}{Dinh Phung}, {and} \bibinfo{person}{Svetha Venkatesh}.} \bibinfo{year}{2008}\natexlab{}.
\newblock \showarticletitle{Sensing and using social context}.
\newblock \bibinfo{journal}{\emph{ACM Transactions on Multimedia Computing, Communications, and Applications (TOMM)}} \bibinfo{volume}{5}, \bibinfo{number}{2} (\bibinfo{year}{2008}), \bibinfo{pages}{1--27}.
\newblock


\bibitem[Ali et~al\mbox{.}(2011)]%
        {ali2011social}
\bibfield{author}{\bibinfo{person}{Raian Ali}, \bibinfo{person}{Carlos Solis}, \bibinfo{person}{Mazeiar Salehie}, \bibinfo{person}{Inah Omoronyia}, \bibinfo{person}{Bashar Nuseibeh}, {and} \bibinfo{person}{Walid Maalej}.} \bibinfo{year}{2011}\natexlab{}.
\newblock \showarticletitle{Social sensing: when users become monitors}.
\newblock \bibinfo{journal}{\emph{Proceedings of the 19th ACM SIGSOFT symposium and the 13th European conference on Foundations of software engineering}}, \bibinfo{pages}{476--479}.
\newblock


\bibitem[Amarasinghe et~al\mbox{.}(2023)]%
        {amarasinghe2023multimodal}
\bibfield{author}{\bibinfo{person}{Yasith Amarasinghe}, \bibinfo{person}{Darshana Sandaruwan}, \bibinfo{person}{Thilina Madusanka}, \bibinfo{person}{Indika Perera}, {and} \bibinfo{person}{Lakmal Meegahapola}.} \bibinfo{year}{2023}\natexlab{}.
\newblock \showarticletitle{Multimodal Earable Sensing for Human Energy Expenditure Estimation}.
\newblock \bibinfo{journal}{\emph{arXiv preprint arXiv:2305.00517}} (\bibinfo{year}{2023}).
\newblock


\bibitem[Assi et~al\mbox{.}(2023)]%
        {assimeegahapola2023complex}
\bibfield{author}{\bibinfo{person}{Karim Assi}, \bibinfo{person}{Lakmal Meegahapola}, \bibinfo{person}{William Droz}, \bibinfo{person}{Peter Kun}, \bibinfo{person}{Amalia de G\"{o}tzen}, \bibinfo{person}{Chaitanya Nutakki}, \bibinfo{person}{Shyam Diwakar}, \bibinfo{person}{Salvador~Ruiz Correa}, \bibinfo{person}{Donglei Song}, \bibinfo{person}{Hao Xu}, \bibinfo{person}{Miriam Bidoglia}, \bibinfo{person}{George Gaskell}, \bibinfo{person}{Altangerel Chagnaa}, \bibinfo{person}{Amarsanaa Ganbold}, \bibinfo{person}{Tsolmon Zundui}, \bibinfo{person}{Carlo Caprini}, \bibinfo{person}{Daniele Miorandi}, \bibinfo{person}{Alethia Hume}, \bibinfo{person}{Jose~Luis Zarza}, \bibinfo{person}{Luca Cernuzzi}, \bibinfo{person}{Ivano Bison}, \bibinfo{person}{Marcelo~Rodas Britez}, \bibinfo{person}{Matteo Busso}, \bibinfo{person}{Ronald Chenu-Abente}, \bibinfo{person}{Can G\"{u}nel}, \bibinfo{person}{Fausto Giunchiglia}, \bibinfo{person}{Laura Schelenz}, {and} \bibinfo{person}{Daniel Gatica-Perez}.}
  \bibinfo{year}{2023}\natexlab{}.
\newblock \showarticletitle{Complex Daily Activities, Country-Level Diversity, and Smartphone Sensing: A Study in Denmark, Italy, Mongolia, Paraguay, and UK}.
\newblock \bibinfo{journal}{\emph{Proceedings of the 2023 CHI Conference on Human Factors in Computing Systems}} (\bibinfo{year}{2023}).
\newblock
\urldef\tempurl%
\url{https://doi.org/10.1145/3569483}
\showDOI{\tempurl}


\bibitem[Bae et~al\mbox{.}(2017)]%
        {bae2017detecting}
\bibfield{author}{\bibinfo{person}{Sangwon Bae}, \bibinfo{person}{Denzil Ferreira}, \bibinfo{person}{Brian Suffoletto}, \bibinfo{person}{Juan~C. Puyana}, \bibinfo{person}{Ryan Kurtz}, \bibinfo{person}{Tammy Chung}, {and} \bibinfo{person}{Anind~K. Dey}.} \bibinfo{year}{2017}\natexlab{}.
\newblock \showarticletitle{Detecting Drinking Episodes in Young Adults Using Smartphone-Based Sensors}.
\newblock \bibinfo{journal}{\emph{Proc. ACM Interact. Mob. Wearable Ubiquitous Technol.}} \bibinfo{volume}{1}, \bibinfo{number}{2}, Article \bibinfo{articleno}{5} (\bibinfo{date}{jun} \bibinfo{year}{2017}), \bibinfo{numpages}{36}~pages.
\newblock
\urldef\tempurl%
\url{https://doi.org/10.1145/3090051}
\showDOI{\tempurl}


\bibitem[Bangamuarachchi et~al\mbox{.}(2022)]%
        {bangamuarachchi2022sensing}
\bibfield{author}{\bibinfo{person}{Wageesha Bangamuarachchi}, \bibinfo{person}{Anju Chamantha}, \bibinfo{person}{Lakmal Meegahapola}, \bibinfo{person}{Salvador Ruiz-Correa}, \bibinfo{person}{Indika Perera}, {and} \bibinfo{person}{Daniel Gatica-Perez}.} \bibinfo{year}{2022}\natexlab{}.
\newblock \showarticletitle{Sensing Eating Events in Context: A Smartphone-Only Approach}.
\newblock \bibinfo{journal}{\emph{IEEE Access}}  \bibinfo{volume}{10} (\bibinfo{year}{2022}), \bibinfo{pages}{61249--61264}.
\newblock
\urldef\tempurl%
\url{https://doi.org/10.1109/ACCESS.2022.3179702}
\showDOI{\tempurl}


\bibitem[Barreto et~al\mbox{.}(2021)]%
        {barreto2021loneliness}
\bibfield{author}{\bibinfo{person}{Manuela Barreto}, \bibinfo{person}{Christina Victor}, \bibinfo{person}{Claudia Hammond}, \bibinfo{person}{Alice Eccles}, \bibinfo{person}{Matt~T Richins}, {and} \bibinfo{person}{Pamela Qualter}.} \bibinfo{year}{2021}\natexlab{}.
\newblock \showarticletitle{Loneliness around the world: Age, gender, and cultural differences in loneliness}.
\newblock \bibinfo{journal}{\emph{Personality and Individual Differences}}  \bibinfo{volume}{169} (\bibinfo{year}{2021}), \bibinfo{pages}{110066}.
\newblock


\bibitem[Baumeister and Leary(1995)]%
        {baumeister1995need}
\bibfield{author}{\bibinfo{person}{Roy~F Baumeister} {and} \bibinfo{person}{Mark~R Leary}.} \bibinfo{year}{1995}\natexlab{}.
\newblock \showarticletitle{The need to belong: desire for interpersonal attachments as a fundamental human motivation.}
\newblock \bibinfo{journal}{\emph{Psychological bulletin}} \bibinfo{volume}{117}, \bibinfo{number}{3} (\bibinfo{year}{1995}), \bibinfo{pages}{497}.
\newblock


\bibitem[Ben-Zeev et~al\mbox{.}(2015)]%
        {ben2015next}
\bibfield{author}{\bibinfo{person}{Dror Ben-Zeev}, \bibinfo{person}{Emily~A Scherer}, \bibinfo{person}{Rui Wang}, \bibinfo{person}{Haiyi Xie}, {and} \bibinfo{person}{Andrew~T Campbell}.} \bibinfo{year}{2015}\natexlab{}.
\newblock \showarticletitle{Next-generation psychiatric assessment: Using smartphone sensors to monitor behavior and mental health.}
\newblock \bibinfo{journal}{\emph{Psychiatric rehabilitation journal}} \bibinfo{volume}{38}, \bibinfo{number}{3} (\bibinfo{year}{2015}), \bibinfo{pages}{218}.
\newblock


\bibitem[Berke et~al\mbox{.}(2011)]%
        {berke2011objective}
\bibfield{author}{\bibinfo{person}{Ethan~M Berke}, \bibinfo{person}{Tanzeem Choudhury}, \bibinfo{person}{Shahid Ali}, {and} \bibinfo{person}{Mashfiqui Rabbi}.} \bibinfo{year}{2011}\natexlab{}.
\newblock \showarticletitle{Objective measurement of sociability and activity: mobile sensing in the community}.
\newblock \bibinfo{journal}{\emph{The Annals of Family Medicine}} \bibinfo{volume}{9}, \bibinfo{number}{4} (\bibinfo{year}{2011}), \bibinfo{pages}{344--350}.
\newblock


\bibitem[Bhattacharjee et~al\mbox{.}(2023)]%
        {bhattacharjee2023integrating}
\bibfield{author}{\bibinfo{person}{Ananya Bhattacharjee}, \bibinfo{person}{Dana Kulzhabayeva}, \bibinfo{person}{Mohi Reza}, \bibinfo{person}{Harsh Kumar}, \bibinfo{person}{Eunchae Seong}, \bibinfo{person}{Xuening Wu}, \bibinfo{person}{Mohammad~Rashidujjaman Rifat}, \bibinfo{person}{Robert Bowman}, \bibinfo{person}{Rachel Kornfield}, \bibinfo{person}{Alex Mariakakis}, {et~al\mbox{.}}} \bibinfo{year}{2023}\natexlab{}.
\newblock \showarticletitle{Integrating Individual and Social Contexts into Self-Reflection Technologies}.
\newblock \bibinfo{journal}{\emph{Extended Abstracts of the 2023 CHI Conference on Human Factors in Computing Systems}}, \bibinfo{pages}{1--6}.
\newblock


\bibitem[Biel et~al\mbox{.}(2018)]%
        {biel2018bites}
\bibfield{author}{\bibinfo{person}{Joan-Isaac Biel}, \bibinfo{person}{Nathalie Martin}, \bibinfo{person}{David Labbe}, {and} \bibinfo{person}{Daniel Gatica-Perez}.} \bibinfo{year}{2018}\natexlab{}.
\newblock \showarticletitle{Bites ‘n’bits: Inferring eating behavior from contextual mobile data}.
\newblock \bibinfo{journal}{\emph{Proceedings of the ACM on Interactive, Mobile, Wearable and Ubiquitous Technologies}} \bibinfo{volume}{1}, \bibinfo{number}{4} (\bibinfo{year}{2018}), \bibinfo{pages}{1--33}.
\newblock


\bibitem[Blakemore(2012)]%
        {blakemore2012development}
\bibfield{author}{\bibinfo{person}{Sarah-Jayne Blakemore}.} \bibinfo{year}{2012}\natexlab{}.
\newblock \showarticletitle{Development of the social brain in adolescence}.
\newblock \bibinfo{journal}{\emph{Journal of the Royal Society of Medicine}} \bibinfo{volume}{105}, \bibinfo{number}{3} (\bibinfo{year}{2012}), \bibinfo{pages}{111--116}.
\newblock


\bibitem[Bolukbasi et~al\mbox{.}(2016)]%
        {bolukbasi2016man}
\bibfield{author}{\bibinfo{person}{Tolga Bolukbasi}, \bibinfo{person}{Kai-Wei Chang}, \bibinfo{person}{James~Y Zou}, \bibinfo{person}{Venkatesh Saligrama}, {and} \bibinfo{person}{Adam~T Kalai}.} \bibinfo{year}{2016}\natexlab{}.
\newblock \showarticletitle{Man is to computer programmer as woman is to homemaker? debiasing word embeddings}.
\newblock \bibinfo{journal}{\emph{Advances in neural information processing systems}}  \bibinfo{volume}{29} (\bibinfo{year}{2016}).
\newblock


\bibitem[Bouton-Bessac et~al\mbox{.}(2022)]%
        {bouton2022your}
\bibfield{author}{\bibinfo{person}{Emma Bouton-Bessac}, \bibinfo{person}{Lakmal Meegahapola}, {and} \bibinfo{person}{Daniel Gatica-Perez}.} \bibinfo{year}{2022}\natexlab{}.
\newblock \showarticletitle{Your Day in Your Pocket: Complex Activity Recognition from Smartphone Accelerometers}. In \bibinfo{booktitle}{\emph{International Conference on Pervasive Computing Technologies for Healthcare}}. Springer, \bibinfo{pages}{247--258}.
\newblock


\bibitem[Broom(2021)]%
        {broom2021home}
\bibfield{author}{\bibinfo{person}{Douglas Broom}.} \bibinfo{year}{2021}\natexlab{}.
\newblock \showarticletitle{Home or office? Survey shows opinions about work after COVID-19}.
\newblock \bibinfo{journal}{\emph{World Economic Forum. July}}  \bibinfo{volume}{21}.
\newblock


\bibitem[Buolamwini and Gebru(2018)]%
        {buolamwini2018gender}
\bibfield{author}{\bibinfo{person}{Joy Buolamwini} {and} \bibinfo{person}{Timnit Gebru}.} \bibinfo{year}{2018}\natexlab{}.
\newblock \showarticletitle{Gender shades: Intersectional accuracy disparities in commercial gender classification}.
\newblock \bibinfo{journal}{\emph{Conference on fairness, accountability and transparency}}, \bibinfo{pages}{77--91}.
\newblock


\bibitem[Burns et~al\mbox{.}(2011)]%
        {burns2011harnessing}
\bibfield{author}{\bibinfo{person}{Michelle~Nicole Burns}, \bibinfo{person}{Mark Begale}, \bibinfo{person}{Jennifer Duffecy}, \bibinfo{person}{Darren Gergle}, \bibinfo{person}{Chris~J Karr}, \bibinfo{person}{Emily Giangrande}, {and} \bibinfo{person}{David~C Mohr}.} \bibinfo{year}{2011}\natexlab{}.
\newblock \showarticletitle{Harnessing context sensing to develop a mobile intervention for depression}.
\newblock \bibinfo{journal}{\emph{Journal of medical Internet research}} \bibinfo{volume}{13}, \bibinfo{number}{3} (\bibinfo{year}{2011}), \bibinfo{pages}{e55}.
\newblock


\bibitem[Can et~al\mbox{.}(2019)]%
        {can2019stress}
\bibfield{author}{\bibinfo{person}{Yekta~Said Can}, \bibinfo{person}{Bert Arnrich}, {and} \bibinfo{person}{Cem Ersoy}.} \bibinfo{year}{2019}\natexlab{}.
\newblock \showarticletitle{Stress detection in daily life scenarios using smart phones and wearable sensors: A survey}.
\newblock \bibinfo{journal}{\emph{Journal of biomedical informatics}}  \bibinfo{volume}{92} (\bibinfo{year}{2019}), \bibinfo{pages}{103139}.
\newblock


\bibitem[Chang et~al\mbox{.}(2020)]%
        {chang2020systematic}
\bibfield{author}{\bibinfo{person}{Youngjae Chang}, \bibinfo{person}{Akhil Mathur}, \bibinfo{person}{Anton Isopoussu}, \bibinfo{person}{Junehwa Song}, {and} \bibinfo{person}{Fahim Kawsar}.} \bibinfo{year}{2020}\natexlab{}.
\newblock \showarticletitle{A systematic study of unsupervised domain adaptation for robust human-activity recognition}.
\newblock \bibinfo{journal}{\emph{Proceedings of the ACM on Interactive, Mobile, Wearable and Ubiquitous Technologies}} \bibinfo{volume}{4}, \bibinfo{number}{1} (\bibinfo{year}{2020}), \bibinfo{pages}{1--30}.
\newblock


\bibitem[Chawla et~al\mbox{.}(2002)]%
        {chawla2002smote}
\bibfield{author}{\bibinfo{person}{Nitesh~V Chawla}, \bibinfo{person}{Kevin~W Bowyer}, \bibinfo{person}{Lawrence~O Hall}, {and} \bibinfo{person}{W~Philip Kegelmeyer}.} \bibinfo{year}{2002}\natexlab{}.
\newblock \showarticletitle{SMOTE: synthetic minority over-sampling technique}.
\newblock \bibinfo{journal}{\emph{Journal of artificial intelligence research}}  \bibinfo{volume}{16} (\bibinfo{year}{2002}), \bibinfo{pages}{321--357}.
\newblock


\bibitem[Chen and Kotz(2000)]%
        {chen2000survey}
\bibfield{author}{\bibinfo{person}{Guanling Chen} {and} \bibinfo{person}{David Kotz}.} \bibinfo{year}{2000}\natexlab{}.
\newblock \showarticletitle{A survey of context-aware mobile computing research}.
\newblock  (\bibinfo{year}{2000}).
\newblock


\bibitem[Chen and Guestrin(2016)]%
        {Chen:2016:XST:2939672.2939785}
\bibfield{author}{\bibinfo{person}{Tianqi Chen} {and} \bibinfo{person}{Carlos Guestrin}.} \bibinfo{year}{2016}\natexlab{}.
\newblock \showarticletitle{{XGBoost}: A Scalable Tree Boosting System}.
\newblock \bibinfo{journal}{\emph{Proceedings of the 22nd ACM SIGKDD International Conference on Knowledge Discovery and Data Mining}}, \bibinfo{pages}{785--794}.
\newblock
\showISBNx{978-1-4503-4232-2}
\urldef\tempurl%
\url{https://doi.org/10.1145/2939672.2939785}
\showDOI{\tempurl}


\bibitem[Cheung and Lim(2021)]%
        {cheung2021systematic}
\bibfield{author}{\bibinfo{person}{Hoi~Shan Cheung} {and} \bibinfo{person}{Elinor Lim}.} \bibinfo{year}{2021}\natexlab{}.
\newblock \showarticletitle{A systematic review of parenting in Singapore: Insights to the culture-specific functions of styles and practices}.
\newblock  (\bibinfo{year}{2021}).
\newblock


\bibitem[Consolvo et~al\mbox{.}(2008)]%
        {consolvo2008activity}
\bibfield{author}{\bibinfo{person}{Sunny Consolvo}, \bibinfo{person}{David~W McDonald}, \bibinfo{person}{Tammy Toscos}, \bibinfo{person}{Mike~Y Chen}, \bibinfo{person}{Jon Froehlich}, \bibinfo{person}{Beverly Harrison}, \bibinfo{person}{Predrag Klasnja}, \bibinfo{person}{Anthony LaMarca}, \bibinfo{person}{Louis LeGrand}, \bibinfo{person}{Ryan Libby}, {et~al\mbox{.}}} \bibinfo{year}{2008}\natexlab{}.
\newblock \showarticletitle{Activity sensing in the wild: a field trial of ubifit garden}.
\newblock \bibinfo{journal}{\emph{Proceedings of the SIGCHI conference on human factors in computing systems}}, \bibinfo{pages}{1797--1806}.
\newblock


\bibitem[Deng et~al\mbox{.}(2009)]%
        {deng2009imagenet}
\bibfield{author}{\bibinfo{person}{Jia Deng}, \bibinfo{person}{Wei Dong}, \bibinfo{person}{Richard Socher}, \bibinfo{person}{Li-Jia Li}, \bibinfo{person}{Kai Li}, {and} \bibinfo{person}{Li Fei-Fei}.} \bibinfo{year}{2009}\natexlab{}.
\newblock \showarticletitle{Imagenet: A large-scale hierarchical image database}.
\newblock \bibinfo{journal}{\emph{2009 IEEE conference on computer vision and pattern recognition}}, \bibinfo{pages}{248--255}.
\newblock


\bibitem[Developers(2021)]%
        {googleActivityApi}
\bibfield{author}{\bibinfo{person}{Google Developers}.} \bibinfo{year}{2021}\natexlab{}.
\newblock \bibinfo{title}{Activity Recognition API}.
\newblock \bibinfo{howpublished}{\url{https://developers.google.com/location-context/activity-recognition}}.
\newblock
\newblock
\shownote{Accessed: 2022-02-15}.


\bibitem[Devlic et~al\mbox{.}(2009)]%
        {devlic2009context}
\bibfield{author}{\bibinfo{person}{Alisa Devlic}, \bibinfo{person}{Roland Reichle}, \bibinfo{person}{Michal Wagner}, \bibinfo{person}{Manuele~Kirsch Pinheiro}, \bibinfo{person}{Yves Vanrompay}, \bibinfo{person}{Yolande Berbers}, {and} \bibinfo{person}{Massimo Valla}.} \bibinfo{year}{2009}\natexlab{}.
\newblock \showarticletitle{Context inference of users' social relationships and distributed policy management}.
\newblock \bibinfo{journal}{\emph{2009 IEEE international conference on pervasive computing and communications}}, \bibinfo{pages}{1--8}.
\newblock


\bibitem[Dey(2001)]%
        {dey2001understanding}
\bibfield{author}{\bibinfo{person}{Anind~K Dey}.} \bibinfo{year}{2001}\natexlab{}.
\newblock \showarticletitle{Understanding and using context}.
\newblock \bibinfo{journal}{\emph{Personal and ubiquitous computing}} \bibinfo{volume}{5}, \bibinfo{number}{1} (\bibinfo{year}{2001}), \bibinfo{pages}{4--7}.
\newblock


\bibitem[Dey et~al\mbox{.}(2001)]%
        {dey2001conceptual}
\bibfield{author}{\bibinfo{person}{Anind~K Dey}, \bibinfo{person}{Gregory~D Abowd}, {and} \bibinfo{person}{Daniel Salber}.} \bibinfo{year}{2001}\natexlab{}.
\newblock \showarticletitle{A conceptual framework and a toolkit for supporting the rapid prototyping of context-aware applications}.
\newblock \bibinfo{journal}{\emph{Human--Computer Interaction}} \bibinfo{volume}{16}, \bibinfo{number}{2-4} (\bibinfo{year}{2001}), \bibinfo{pages}{97--166}.
\newblock


\bibitem[Dwork et~al\mbox{.}(2012)]%
        {dwork2012fairness}
\bibfield{author}{\bibinfo{person}{Cynthia Dwork}, \bibinfo{person}{Moritz Hardt}, \bibinfo{person}{Toniann Pitassi}, \bibinfo{person}{Omer Reingold}, {and} \bibinfo{person}{Richard Zemel}.} \bibinfo{year}{2012}\natexlab{}.
\newblock \showarticletitle{Fairness through awareness}.
\newblock \bibinfo{journal}{\emph{Proceedings of the 3rd innovations in theoretical computer science conference}}, \bibinfo{pages}{214--226}.
\newblock


\bibitem[Fried et~al\mbox{.}(2022)]%
        {fried2022mental}
\bibfield{author}{\bibinfo{person}{Eiko~I Fried}, \bibinfo{person}{Faidra Papanikolaou}, {and} \bibinfo{person}{Sacha Epskamp}.} \bibinfo{year}{2022}\natexlab{}.
\newblock \showarticletitle{Mental health and social contact during the COVID-19 pandemic: an ecological momentary assessment study}.
\newblock \bibinfo{journal}{\emph{Clinical Psychological Science}} \bibinfo{volume}{10}, \bibinfo{number}{2} (\bibinfo{year}{2022}), \bibinfo{pages}{340--354}.
\newblock


\bibitem[Fry and Parker(2021)]%
        {fry2021rising}
\bibfield{author}{\bibinfo{person}{Richard Fry} {and} \bibinfo{person}{Kim Parker}.} \bibinfo{year}{2021}\natexlab{}.
\newblock \showarticletitle{Rising share of US adults are living without a spouse or partner}.
\newblock  (\bibinfo{year}{2021}).
\newblock


\bibitem[Fuligni and Eccles(1993)]%
        {fuligni1993perceived}
\bibfield{author}{\bibinfo{person}{Andrew~J Fuligni} {and} \bibinfo{person}{Jacquelynne~S Eccles}.} \bibinfo{year}{1993}\natexlab{}.
\newblock \showarticletitle{Perceived parent-child relationships and early adolescents' orientation toward peers.}
\newblock \bibinfo{journal}{\emph{Developmental psychology}} \bibinfo{volume}{29}, \bibinfo{number}{4} (\bibinfo{year}{1993}), \bibinfo{pages}{622}.
\newblock


\bibitem[Giunchiglia et~al\mbox{.}(2022)]%
        {giunchiglia2022worldwide}
\bibfield{author}{\bibinfo{person}{Fausto Giunchiglia}, \bibinfo{person}{Ivano Bison}, \bibinfo{person}{Matteo Busso}, \bibinfo{person}{Ronald Chenu-Abente}, \bibinfo{person}{Marcelo Rodas}, \bibinfo{person}{Mattia Zeni}, \bibinfo{person}{Can Gunel}, \bibinfo{person}{Giuseppe Veltri}, \bibinfo{person}{Amalia De~G{\"o}tzen}, \bibinfo{person}{Peter Kun}, {et~al\mbox{.}}} \bibinfo{year}{2022}\natexlab{}.
\newblock \bibinfo{title}{A worldwide diversity pilot on daily routines and social practices (2020-2021). University of Trento Technical Report-DataScientia dataset descriptors}.
\newblock
\newblock


\bibitem[Gong et~al\mbox{.}(2019)]%
        {gong2019diversity}
\bibfield{author}{\bibinfo{person}{Zhiqiang Gong}, \bibinfo{person}{Ping Zhong}, {and} \bibinfo{person}{Weidong Hu}.} \bibinfo{year}{2019}\natexlab{}.
\newblock \showarticletitle{Diversity in machine learning}.
\newblock \bibinfo{journal}{\emph{IEEE Access}}  \bibinfo{volume}{7} (\bibinfo{year}{2019}), \bibinfo{pages}{64323--64350}.
\newblock


\bibitem[Goody(1996)]%
        {goody1996comparing}
\bibfield{author}{\bibinfo{person}{Jack Goody}.} \bibinfo{year}{1996}\natexlab{}.
\newblock \showarticletitle{Comparing family systems in Europe and Asia: Are there different sets of rules?}
\newblock \bibinfo{journal}{\emph{Population and Development Review}} (\bibinfo{year}{1996}), \bibinfo{pages}{1--20}.
\newblock


\bibitem[Heron(1970)]%
        {heron1970phenomenology}
\bibfield{author}{\bibinfo{person}{John Heron}.} \bibinfo{year}{1970}\natexlab{}.
\newblock \showarticletitle{The phenomenology of social encounter: The gaze}.
\newblock \bibinfo{journal}{\emph{Philosophy and Phenomenological Research}} \bibinfo{volume}{31}, \bibinfo{number}{2} (\bibinfo{year}{1970}), \bibinfo{pages}{243--264}.
\newblock


\bibitem[Holt-Lunstad et~al\mbox{.}(2015)]%
        {holt2015loneliness}
\bibfield{author}{\bibinfo{person}{Julianne Holt-Lunstad}, \bibinfo{person}{Timothy~B Smith}, \bibinfo{person}{Mark Baker}, \bibinfo{person}{Tyler Harris}, {and} \bibinfo{person}{David Stephenson}.} \bibinfo{year}{2015}\natexlab{}.
\newblock \showarticletitle{Loneliness and social isolation as risk factors for mortality: a meta-analytic review}.
\newblock \bibinfo{journal}{\emph{Perspectives on psychological science}} \bibinfo{volume}{10}, \bibinfo{number}{2} (\bibinfo{year}{2015}), \bibinfo{pages}{227--237}.
\newblock


\bibitem[Hu et~al\mbox{.}(2017)]%
        {hu2017elderly}
\bibfield{author}{\bibinfo{person}{Rui Hu}, \bibinfo{person}{Hieu Pham}, \bibinfo{person}{Philipp Buluschek}, {and} \bibinfo{person}{Daniel Gatica-Perez}.} \bibinfo{year}{2017}\natexlab{}.
\newblock \showarticletitle{Elderly people living alone: Detecting home visits with ambient and wearable sensing}.
\newblock \bibinfo{journal}{\emph{Proceedings of the 2nd International Workshop on Multimedia for Personal Health and Health Care}}, \bibinfo{pages}{85--88}.
\newblock


\bibitem[Kammoun et~al\mbox{.}(2023)]%
        {kammoun2023understanding}
\bibfield{author}{\bibinfo{person}{Nathan Kammoun}, \bibinfo{person}{Lakmal Meegahapola}, {and} \bibinfo{person}{Daniel Gatica-Perez}.} \bibinfo{year}{2023}\natexlab{}.
\newblock \showarticletitle{Understanding the Social Context of Eating with Multimodal Smartphone Sensing: The Role of Country Diversity}.
\newblock \bibinfo{journal}{\emph{Proceedings of the 25th International Conference on Multimodal Interaction (ICMI '23)}} (\bibinfo{year}{2023}).
\newblock


\bibitem[Kapoor and Narayanan(2022)]%
        {kapoor2022leakage}
\bibfield{author}{\bibinfo{person}{Sayash Kapoor} {and} \bibinfo{person}{Arvind Narayanan}.} \bibinfo{year}{2022}\natexlab{}.
\newblock \showarticletitle{Leakage and the reproducibility crisis in ML-based science}.
\newblock \bibinfo{journal}{\emph{arXiv preprint arXiv:2207.07048}} (\bibinfo{year}{2022}).
\newblock


\bibitem[Khwaja et~al\mbox{.}(2019)]%
        {khwaja2019modeling}
\bibfield{author}{\bibinfo{person}{Mohammed Khwaja}, \bibinfo{person}{Sumer~S Vaid}, \bibinfo{person}{Sara Zannone}, \bibinfo{person}{Gabriella~M Harari}, \bibinfo{person}{A~Aldo Faisal}, {and} \bibinfo{person}{Aleksandar Matic}.} \bibinfo{year}{2019}\natexlab{}.
\newblock \showarticletitle{Modeling personality vs. modeling personalidad: In-the-wild mobile data analysis in five countries suggests cultural impact on personality models}.
\newblock \bibinfo{journal}{\emph{Proceedings of the ACM on Interactive, Mobile, Wearable and Ubiquitous Technologies}} \bibinfo{volume}{3}, \bibinfo{number}{3} (\bibinfo{year}{2019}), \bibinfo{pages}{1--24}.
\newblock


\bibitem[Kim et~al\mbox{.}(2015)]%
        {kim2015relationships}
\bibfield{author}{\bibinfo{person}{Kyungmin Kim}, \bibinfo{person}{Yen-Pi Cheng}, \bibinfo{person}{Steven~H Zarit}, {and} \bibinfo{person}{Karen~L Fingerman}.} \bibinfo{year}{2015}\natexlab{}.
\newblock \showarticletitle{Relationships between adults and parents in Asia}.
\newblock \bibinfo{journal}{\emph{Successful aging: Asian perspectives}} (\bibinfo{year}{2015}), \bibinfo{pages}{101--122}.
\newblock


\bibitem[Krasin et~al\mbox{.}(2017)]%
        {krasin2017openimages}
\bibfield{author}{\bibinfo{person}{Ivan Krasin}, \bibinfo{person}{Tom Duerig}, \bibinfo{person}{Neil Alldrin}, \bibinfo{person}{Vittorio Ferrari}, \bibinfo{person}{Sami Abu-El-Haija}, \bibinfo{person}{Alina Kuznetsova}, \bibinfo{person}{Hassan Rom}, \bibinfo{person}{Jasper Uijlings}, \bibinfo{person}{Stefan Popov}, \bibinfo{person}{Andreas Veit}, {et~al\mbox{.}}} \bibinfo{year}{2017}\natexlab{}.
\newblock \showarticletitle{Openimages: A public dataset for large-scale multi-label and multi-class image classification}.
\newblock \bibinfo{journal}{\emph{Dataset available from https://github. com/openimages}} \bibinfo{volume}{2}, \bibinfo{number}{3} (\bibinfo{year}{2017}), \bibinfo{pages}{18}.
\newblock


\bibitem[Liang and Cao(2015)]%
        {liang2015social}
\bibfield{author}{\bibinfo{person}{Guanqing Liang} {and} \bibinfo{person}{Jiannong Cao}.} \bibinfo{year}{2015}\natexlab{}.
\newblock \showarticletitle{Social context-aware middleware: A survey}.
\newblock \bibinfo{journal}{\emph{Pervasive and mobile computing}}  \bibinfo{volume}{17} (\bibinfo{year}{2015}), \bibinfo{pages}{207--219}.
\newblock


\bibitem[LiKamWa et~al\mbox{.}(2013)]%
        {likamwa2013moodscope}
\bibfield{author}{\bibinfo{person}{Robert LiKamWa}, \bibinfo{person}{Yunxin Liu}, \bibinfo{person}{Nicholas~D Lane}, {and} \bibinfo{person}{Lin Zhong}.} \bibinfo{year}{2013}\natexlab{}.
\newblock \showarticletitle{Moodscope: Building a mood sensor from smartphone usage patterns}.
\newblock \bibinfo{journal}{\emph{Proceeding of the 11th annual international conference on Mobile systems, applications, and services}}, \bibinfo{pages}{389--402}.
\newblock


\bibitem[Little and Rubin(2019)]%
        {little2019statistical}
\bibfield{author}{\bibinfo{person}{Roderick~JA Little} {and} \bibinfo{person}{Donald~B Rubin}.} \bibinfo{year}{2019}\natexlab{}.
\newblock \bibinfo{booktitle}{\emph{Statistical analysis with missing data}}. Vol.~\bibinfo{volume}{793}.
\newblock \bibinfo{publisher}{John Wiley \& Sons}.
\newblock


\bibitem[Lockhart et~al\mbox{.}(2012)]%
        {lockhart2012applications}
\bibfield{author}{\bibinfo{person}{Jeffrey~W Lockhart}, \bibinfo{person}{Tony Pulickal}, {and} \bibinfo{person}{Gary~M Weiss}.} \bibinfo{year}{2012}\natexlab{}.
\newblock \showarticletitle{Applications of mobile activity recognition}.
\newblock \bibinfo{journal}{\emph{Proceedings of the 2012 ACM Conference on Ubiquitous Computing}}, \bibinfo{pages}{1054--1058}.
\newblock


\bibitem[Lund et~al\mbox{.}(2021)]%
        {lund2021future}
\bibfield{author}{\bibinfo{person}{Susan Lund}, \bibinfo{person}{Anu Madgavkar}, \bibinfo{person}{James Manyika}, \bibinfo{person}{Sven Smit}, \bibinfo{person}{Kweilin Ellingrud}, \bibinfo{person}{Mary Meaney}, {and} \bibinfo{person}{Olivia Robinson}.} \bibinfo{year}{2021}\natexlab{}.
\newblock \showarticletitle{The future of work after COVID-19}.
\newblock \bibinfo{journal}{\emph{McKinsey global institute}}  \bibinfo{volume}{18} (\bibinfo{year}{2021}).
\newblock


\bibitem[Mass(2021)]%
        {mass2021work}
\bibfield{author}{\bibinfo{person}{S Mass}.} \bibinfo{year}{2021}\natexlab{}.
\newblock \showarticletitle{Work from home likely to remain elevated post pandemic}.
\newblock \bibinfo{journal}{\emph{The Digest}}  \bibinfo{volume}{6} (\bibinfo{year}{2021}).
\newblock


\bibitem[Matthews et~al\mbox{.}(2016)]%
        {matthews2016social}
\bibfield{author}{\bibinfo{person}{Timothy Matthews}, \bibinfo{person}{Andrea Danese}, \bibinfo{person}{Jasmin Wertz}, \bibinfo{person}{Candice~L Odgers}, \bibinfo{person}{Antony Ambler}, \bibinfo{person}{Terrie~E Moffitt}, {and} \bibinfo{person}{Louise Arseneault}.} \bibinfo{year}{2016}\natexlab{}.
\newblock \showarticletitle{Social isolation, loneliness and depression in young adulthood: a behavioural genetic analysis}.
\newblock \bibinfo{journal}{\emph{Social psychiatry and psychiatric epidemiology}} \bibinfo{volume}{51}, \bibinfo{number}{3} (\bibinfo{year}{2016}), \bibinfo{pages}{339--348}.
\newblock


\bibitem[Mayer et~al\mbox{.}(2015)]%
        {mayer2015making}
\bibfield{author}{\bibinfo{person}{Julia~M Mayer}, \bibinfo{person}{Starr~Roxanne Hiltz}, {and} \bibinfo{person}{Quentin Jones}.} \bibinfo{year}{2015}\natexlab{}.
\newblock \showarticletitle{Making social matching context-aware: Design concepts and open challenges}.
\newblock \bibinfo{journal}{\emph{Proceedings of the 33rd Annual ACM Conference on Human Factors in Computing Systems}}, \bibinfo{pages}{545--554}.
\newblock


\bibitem[Meegahapola et~al\mbox{.}(2023)]%
        {meegahapola2023generalization}
\bibfield{author}{\bibinfo{person}{Lakmal Meegahapola}, \bibinfo{person}{William Droz}, \bibinfo{person}{Peter Kun}, \bibinfo{person}{Amalia de G\"{o}tzen}, \bibinfo{person}{Chaitanya Nutakki}, \bibinfo{person}{Shyam Diwakar}, \bibinfo{person}{Salvador~Ruiz Correa}, \bibinfo{person}{Donglei Song}, \bibinfo{person}{Hao Xu}, \bibinfo{person}{Miriam Bidoglia}, \bibinfo{person}{George Gaskell}, \bibinfo{person}{Altangerel Chagnaa}, \bibinfo{person}{Amarsanaa Ganbold}, \bibinfo{person}{Tsolmon Zundui}, \bibinfo{person}{Carlo Caprini}, \bibinfo{person}{Daniele Miorandi}, \bibinfo{person}{Alethia Hume}, \bibinfo{person}{Jose~Luis Zarza}, \bibinfo{person}{Luca Cernuzzi}, \bibinfo{person}{Ivano Bison}, \bibinfo{person}{Marcelo~Rodas Britez}, \bibinfo{person}{Matteo Busso}, \bibinfo{person}{Ronald Chenu-Abente}, \bibinfo{person}{Can G\"{u}nel}, \bibinfo{person}{Fausto Giunchiglia}, \bibinfo{person}{Laura Schelenz}, {and} \bibinfo{person}{Daniel Gatica-Perez}.} \bibinfo{year}{2023}\natexlab{}.
\newblock \showarticletitle{Generalization and Personalization of Mobile Sensing-Based Mood Inference Models: An Analysis of College Students in Eight Countries}.
\newblock \bibinfo{journal}{\emph{Proc. ACM Interact. Mob. Wearable Ubiquitous Technol.}} \bibinfo{volume}{6}, \bibinfo{number}{4}, Article \bibinfo{articleno}{176} (\bibinfo{date}{jan} \bibinfo{year}{2023}), \bibinfo{numpages}{32}~pages.
\newblock
\urldef\tempurl%
\url{https://doi.org/10.1145/3569483}
\showDOI{\tempurl}


\bibitem[Meegahapola and Gatica-Perez(2020)]%
        {meegahapola2020smartphone}
\bibfield{author}{\bibinfo{person}{Lakmal Meegahapola} {and} \bibinfo{person}{Daniel Gatica-Perez}.} \bibinfo{year}{2020}\natexlab{}.
\newblock \showarticletitle{Smartphone Sensing for the Well-Being of Young Adults: A Review}.
\newblock \bibinfo{journal}{\emph{IEEE Access}} (\bibinfo{year}{2020}).
\newblock


\bibitem[Meegahapola et~al\mbox{.}(2021a)]%
        {meegahapola2021examining}
\bibfield{author}{\bibinfo{person}{Lakmal Meegahapola}, \bibinfo{person}{Florian Labhart}, \bibinfo{person}{Thanh-Trung Phan}, {and} \bibinfo{person}{Daniel Gatica-Perez}.} \bibinfo{year}{2021}\natexlab{a}.
\newblock \showarticletitle{Examining the Social Context of Alcohol Drinking in Young Adults with Smartphone Sensing}.
\newblock \bibinfo{journal}{\emph{Proceedings of the ACM on Interactive, Mobile, Wearable and Ubiquitous Technologies}} \bibinfo{volume}{5}, \bibinfo{number}{3} (\bibinfo{year}{2021}), \bibinfo{pages}{1--26}.
\newblock


\bibitem[Meegahapola et~al\mbox{.}(2020a)]%
        {meegahapola2020alone}
\bibfield{author}{\bibinfo{person}{Lakmal Meegahapola}, \bibinfo{person}{Salvador Ruiz-Correa}, {and} \bibinfo{person}{Daniel Gatica-Perez}.} \bibinfo{year}{2020}\natexlab{a}.
\newblock \showarticletitle{Alone or with others? understanding eating episodes of college students with mobile sensing}.
\newblock \bibinfo{journal}{\emph{19th International Conference on Mobile and Ubiquitous Multimedia}}, \bibinfo{pages}{162--166}.
\newblock


\bibitem[Meegahapola et~al\mbox{.}(2020b)]%
        {meegahapola2020protecting}
\bibfield{author}{\bibinfo{person}{Lakmal Meegahapola}, \bibinfo{person}{Salvador Ruiz-Correa}, {and} \bibinfo{person}{Daniel Gatica-Perez}.} \bibinfo{year}{2020}\natexlab{b}.
\newblock \showarticletitle{Protecting mobile food diaries from getting too personal}.
\newblock \bibinfo{journal}{\emph{Proceedings of the 19th International Conference on Mobile and Ubiquitous Multimedia}}, \bibinfo{pages}{212--222}.
\newblock


\bibitem[Meegahapola et~al\mbox{.}(2021b)]%
        {meegahapola2021one}
\bibfield{author}{\bibinfo{person}{Lakmal Meegahapola}, \bibinfo{person}{Salvador Ruiz-Correa}, \bibinfo{person}{Viridiana del~Carmen Robledo-Valero}, \bibinfo{person}{Emilio~Ernesto Hernandez-Huerfano}, \bibinfo{person}{Leonardo Alvarez-Rivera}, \bibinfo{person}{Ronald Chenu-Abente}, {and} \bibinfo{person}{Daniel Gatica-Perez}.} \bibinfo{year}{2021}\natexlab{b}.
\newblock \showarticletitle{One More Bite? Inferring Food Consumption Level of College Students Using Smartphone Sensing and Self-Reports}.
\newblock \bibinfo{journal}{\emph{Proc. ACM Interact. Mob. Wearable Ubiquitous Technol.}} \bibinfo{volume}{5}, \bibinfo{number}{1}, Article \bibinfo{articleno}{26} (\bibinfo{date}{mar} \bibinfo{year}{2021}), \bibinfo{numpages}{28}~pages.
\newblock
\urldef\tempurl%
\url{https://doi.org/10.1145/3448120}
\showDOI{\tempurl}


\bibitem[Nanchen et~al\mbox{.}(2023)]%
        {nanchen2023keep}
\bibfield{author}{\bibinfo{person}{Alexandre Nanchen}, \bibinfo{person}{Lakmal Meegahapola}, \bibinfo{person}{William Droz}, {and} \bibinfo{person}{Daniel Gatica-Perez}.} \bibinfo{year}{2023}\natexlab{}.
\newblock \showarticletitle{Keep Sensors in Check: Disentangling Country-Level Generalization Issues in Mobile Sensor-Based Models with Diversity Scores}.
\newblock \bibinfo{journal}{\emph{Proceedings of the 2023 AAAI/ACM Conference on AI, Ethics, and Society}}, \bibinfo{pages}{217–228}.
\newblock
\showISBNx{9798400702310}
\urldef\tempurl%
\url{https://doi.org/10.1145/3600211.3604688}
\showDOI{\tempurl}


\bibitem[Nepal et~al\mbox{.}(2022)]%
        {nepal2022covid}
\bibfield{author}{\bibinfo{person}{Subigya Nepal}, \bibinfo{person}{Weichen Wang}, \bibinfo{person}{Vlado Vojdanovski}, \bibinfo{person}{Jeremy~F Huckins}, \bibinfo{person}{Alex Dasilva}, \bibinfo{person}{Meghan Meyer}, {and} \bibinfo{person}{Andrew Campbell}.} \bibinfo{year}{2022}\natexlab{}.
\newblock \showarticletitle{COVID student study: A year in the life of college students during the COVID-19 pandemic through the lens of mobile phone sensing}.
\newblock \bibinfo{journal}{\emph{Proceedings of the 2022 CHI Conference on Human Factors in Computing Systems}}, \bibinfo{pages}{1--19}.
\newblock


\bibitem[Ortiz-Ospina et~al\mbox{.}(2020)]%
        {ortiz2020time}
\bibfield{author}{\bibinfo{person}{Esteban Ortiz-Ospina}, \bibinfo{person}{Charlie Giattino}, {and} \bibinfo{person}{Max Roser}.} \bibinfo{year}{2020}\natexlab{}.
\newblock \showarticletitle{Time use}.
\newblock \bibinfo{journal}{\emph{Our World in Data}} (\bibinfo{year}{2020}).
\newblock


\bibitem[Pedregosa et~al\mbox{.}(2011)]%
        {pedregosa2011scikit}
\bibfield{author}{\bibinfo{person}{Fabian Pedregosa}, \bibinfo{person}{Ga{\"e}l Varoquaux}, \bibinfo{person}{Alexandre Gramfort}, \bibinfo{person}{Vincent Michel}, \bibinfo{person}{Bertrand Thirion}, \bibinfo{person}{Olivier Grisel}, \bibinfo{person}{Mathieu Blondel}, \bibinfo{person}{Peter Prettenhofer}, \bibinfo{person}{Ron Weiss}, \bibinfo{person}{Vincent Dubourg}, {et~al\mbox{.}}} \bibinfo{year}{2011}\natexlab{}.
\newblock \showarticletitle{Scikit-learn: Machine learning in Python}.
\newblock \bibinfo{journal}{\emph{the Journal of machine Learning research}}  \bibinfo{volume}{12} (\bibinfo{year}{2011}), \bibinfo{pages}{2825--2830}.
\newblock


\bibitem[Phan et~al\mbox{.}(2022)]%
        {phan2022mobile}
\bibfield{author}{\bibinfo{person}{Le~Vy Phan}, \bibinfo{person}{Nick Modersitzki}, \bibinfo{person}{Kim~K Gloystein}, {and} \bibinfo{person}{Sandrine M{\"u}ller}.} \bibinfo{year}{2022}\natexlab{}.
\newblock \bibinfo{title}{Mobile Sensing Around the Globe: Considerations for Cross-Cultural Research}.
\newblock
\newblock
\urldef\tempurl%
\url{https://doi.org/10.31234/osf.io/q8c7y}
\showDOI{\tempurl}


\bibitem[Pires et~al\mbox{.}(2016)]%
        {pires2016identification}
\bibfield{author}{\bibinfo{person}{Ivan~Miguel Pires}, \bibinfo{person}{Nuno~M Garcia}, \bibinfo{person}{Nuno Pombo}, {and} \bibinfo{person}{Francisco Fl{\'o}rez-Revuelta}.} \bibinfo{year}{2016}\natexlab{}.
\newblock \showarticletitle{Identification of activities of daily living using sensors available in off-the-shelf mobile devices: Research and hypothesis}.
\newblock \bibinfo{journal}{\emph{International Symposium on Ambient Intelligence}}, \bibinfo{pages}{121--130}.
\newblock


\bibitem[Porfirio et~al\mbox{.}(2020)]%
        {porfirio2020transforming}
\bibfield{author}{\bibinfo{person}{David Porfirio}, \bibinfo{person}{Allison Saupp{\'e}}, \bibinfo{person}{Aws Albarghouthi}, {and} \bibinfo{person}{Bilge Mutlu}.} \bibinfo{year}{2020}\natexlab{}.
\newblock \showarticletitle{Transforming robot programs based on social context}.
\newblock \bibinfo{journal}{\emph{Proceedings of the 2020 CHI conference on human factors in computing systems}}, \bibinfo{pages}{1--12}.
\newblock


\bibitem[Prates et~al\mbox{.}(2020)]%
        {prates2020assessing}
\bibfield{author}{\bibinfo{person}{Marcelo~OR Prates}, \bibinfo{person}{Pedro~H Avelar}, {and} \bibinfo{person}{Lu{\'\i}s~C Lamb}.} \bibinfo{year}{2020}\natexlab{}.
\newblock \showarticletitle{Assessing gender bias in machine translation: a case study with google translate}.
\newblock \bibinfo{journal}{\emph{Neural Computing and Applications}} \bibinfo{volume}{32}, \bibinfo{number}{10} (\bibinfo{year}{2020}), \bibinfo{pages}{6363--6381}.
\newblock


\bibitem[Pudil et~al\mbox{.}(1994)]%
        {pudil1994floating}
\bibfield{author}{\bibinfo{person}{Pavel Pudil}, \bibinfo{person}{Jana Novovi{\v{c}}ov{\'a}}, {and} \bibinfo{person}{Josef Kittler}.} \bibinfo{year}{1994}\natexlab{}.
\newblock \showarticletitle{Floating search methods in feature selection}.
\newblock \bibinfo{journal}{\emph{Pattern recognition letters}} \bibinfo{volume}{15}, \bibinfo{number}{11} (\bibinfo{year}{1994}), \bibinfo{pages}{1119--1125}.
\newblock


\bibitem[Radu et~al\mbox{.}(2018)]%
        {radu2018multimodal}
\bibfield{author}{\bibinfo{person}{Valentin Radu}, \bibinfo{person}{Catherine Tong}, \bibinfo{person}{Sourav Bhattacharya}, \bibinfo{person}{Nicholas~D Lane}, \bibinfo{person}{Cecilia Mascolo}, \bibinfo{person}{Mahesh~K Marina}, {and} \bibinfo{person}{Fahim Kawsar}.} \bibinfo{year}{2018}\natexlab{}.
\newblock \showarticletitle{Multimodal deep learning for activity and context recognition}.
\newblock \bibinfo{journal}{\emph{Proceedings of the ACM on interactive, mobile, wearable and ubiquitous technologies}} \bibinfo{volume}{1}, \bibinfo{number}{4} (\bibinfo{year}{2018}), \bibinfo{pages}{1--27}.
\newblock


\bibitem[Raento et~al\mbox{.}(2005)]%
        {raento2005contextphone}
\bibfield{author}{\bibinfo{person}{Mika Raento}, \bibinfo{person}{Antti Oulasvirta}, \bibinfo{person}{Renaud Petit}, {and} \bibinfo{person}{Hannu Toivonen}.} \bibinfo{year}{2005}\natexlab{}.
\newblock \showarticletitle{ContextPhone: A prototyping platform for context-aware mobile applications}.
\newblock \bibinfo{journal}{\emph{IEEE pervasive computing}} \bibinfo{volume}{4}, \bibinfo{number}{2} (\bibinfo{year}{2005}), \bibinfo{pages}{51--59}.
\newblock


\bibitem[Raschka et~al\mbox{.}(2020)]%
        {raschka2020machine}
\bibfield{author}{\bibinfo{person}{Sebastian Raschka}, \bibinfo{person}{Joshua Patterson}, {and} \bibinfo{person}{Corey Nolet}.} \bibinfo{year}{2020}\natexlab{}.
\newblock \showarticletitle{Machine Learning in Python: Main developments and technology trends in data science, machine learning, and artificial intelligence}.
\newblock \bibinfo{journal}{\emph{arXiv preprint arXiv:2002.04803}} (\bibinfo{year}{2020}).
\newblock


\bibitem[Santani et~al\mbox{.}(2018)]%
        {santani2018drinksense}
\bibfield{author}{\bibinfo{person}{Darshan Santani}, \bibinfo{person}{Florian Labhart}, \bibinfo{person}{Sara Landolt}, \bibinfo{person}{Emmanuel Kuntsche}, \bibinfo{person}{Daniel Gatica-Perez}, {et~al\mbox{.}}} \bibinfo{year}{2018}\natexlab{}.
\newblock \showarticletitle{DrinkSense: Characterizing youth drinking behavior using smartphones}.
\newblock \bibinfo{journal}{\emph{IEEE Transactions on Mobile Computing}} \bibinfo{volume}{17}, \bibinfo{number}{10} (\bibinfo{year}{2018}), \bibinfo{pages}{2279--2292}.
\newblock


\bibitem[Sawyer et~al\mbox{.}(2012)]%
        {sawyer2012using}
\bibfield{author}{\bibinfo{person}{Blake Sawyer}, \bibinfo{person}{Francis Quek}, \bibinfo{person}{Wai~Choong Wong}, \bibinfo{person}{Mehul Motani}, \bibinfo{person}{Sharon~Lynn Chu Yew~Yee}, {and} \bibinfo{person}{Manuel Perez-Quinones}.} \bibinfo{year}{2012}\natexlab{}.
\newblock \showarticletitle{Using physical-social interactions to support information re-finding}.
\newblock \bibinfo{pages}{885--910}.
\newblock


\bibitem[Schelenz et~al\mbox{.}(2021)]%
        {schelenz2021theory}
\bibfield{author}{\bibinfo{person}{Laura Schelenz}, \bibinfo{person}{Ivano Bison}, \bibinfo{person}{Matteo Busso}, \bibinfo{person}{Amalia De~G{\"o}tzen}, \bibinfo{person}{Daniel Gatica-Perez}, \bibinfo{person}{Fausto Giunchiglia}, \bibinfo{person}{Lakmal Meegahapola}, {and} \bibinfo{person}{Salvador Ruiz-Correa}.} \bibinfo{year}{2021}\natexlab{}.
\newblock \showarticletitle{The Theory, Practice, and Ethical Challenges of Designing a Diversity-Aware Platform for Social Relations}.
\newblock \bibinfo{journal}{\emph{Proceedings of the 2021 AAAI/ACM Conference on AI, Ethics, and Society}}, \bibinfo{pages}{905--915}.
\newblock


\bibitem[Servia-Rodr{\'\i}guez et~al\mbox{.}(2017)]%
        {servia2017mobile}
\bibfield{author}{\bibinfo{person}{Sandra Servia-Rodr{\'\i}guez}, \bibinfo{person}{Kiran~K Rachuri}, \bibinfo{person}{Cecilia Mascolo}, \bibinfo{person}{Peter~J Rentfrow}, \bibinfo{person}{Neal Lathia}, {and} \bibinfo{person}{Gillian~M Sandstrom}.} \bibinfo{year}{2017}\natexlab{}.
\newblock \showarticletitle{Mobile sensing at the service of mental well-being: a large-scale longitudinal study}.
\newblock \bibinfo{journal}{\emph{Proceedings of the 26th International Conference on World Wide Web}}, \bibinfo{pages}{103--112}.
\newblock


\bibitem[Shankar et~al\mbox{.}(2017)]%
        {shankar2017no}
\bibfield{author}{\bibinfo{person}{Shreya Shankar}, \bibinfo{person}{Yoni Halpern}, \bibinfo{person}{Eric Breck}, \bibinfo{person}{James Atwood}, \bibinfo{person}{Jimbo Wilson}, {and} \bibinfo{person}{D Sculley}.} \bibinfo{year}{2017}\natexlab{}.
\newblock \showarticletitle{No classification without representation: Assessing geodiversity issues in open data sets for the developing world}.
\newblock \bibinfo{journal}{\emph{arXiv preprint arXiv:1711.08536}} (\bibinfo{year}{2017}).
\newblock


\bibitem[Sheiner and Grasela(1991)]%
        {sheiner1991introduction}
\bibfield{author}{\bibinfo{person}{Lewis~B Sheiner} {and} \bibinfo{person}{Thaddeus~H Grasela}.} \bibinfo{year}{1991}\natexlab{}.
\newblock \showarticletitle{An introduction to mixed effect modeling: concepts, definitions, and justification}.
\newblock \bibinfo{journal}{\emph{Journal of pharmacokinetics and biopharmaceutics}}  \bibinfo{volume}{19} (\bibinfo{year}{1991}), \bibinfo{pages}{S11--S24}.
\newblock


\bibitem[Solmaz and Wu(2017)]%
        {solmaz2017together}
\bibfield{author}{\bibinfo{person}{G{\"u}rkan Solmaz} {and} \bibinfo{person}{Fang-Jing Wu}.} \bibinfo{year}{2017}\natexlab{}.
\newblock \showarticletitle{Together or alone: Detecting group mobility with wireless fingerprints}.
\newblock \bibinfo{journal}{\emph{2017 IEEE International Conference on Communications (ICC)}}, \bibinfo{pages}{1--7}.
\newblock


\bibitem[Stisen et~al\mbox{.}(2015)]%
        {stisen2015smart}
\bibfield{author}{\bibinfo{person}{Allan Stisen}, \bibinfo{person}{Henrik Blunck}, \bibinfo{person}{Sourav Bhattacharya}, \bibinfo{person}{Thor~Siiger Prentow}, \bibinfo{person}{Mikkel~Baun Kj{\ae}rgaard}, \bibinfo{person}{Anind Dey}, \bibinfo{person}{Tobias Sonne}, {and} \bibinfo{person}{Mads~M{\o}ller Jensen}.} \bibinfo{year}{2015}\natexlab{}.
\newblock \showarticletitle{Smart devices are different: Assessing and mitigatingmobile sensing heterogeneities for activity recognition}.
\newblock \bibinfo{journal}{\emph{Proceedings of the 13th ACM conference on embedded networked sensor systems}}, \bibinfo{pages}{127--140}.
\newblock


\bibitem[Tamminen et~al\mbox{.}(2019)]%
        {tamminen2019living}
\bibfield{author}{\bibinfo{person}{Nina Tamminen}, \bibinfo{person}{Tarja Kettunen}, \bibinfo{person}{Tuija Martelin}, \bibinfo{person}{Jaakko Reinikainen}, {and} \bibinfo{person}{Pia Solin}.} \bibinfo{year}{2019}\natexlab{}.
\newblock \showarticletitle{Living alone and positive mental health: a systematic review}.
\newblock \bibinfo{journal}{\emph{Systematic reviews}} \bibinfo{volume}{8}, \bibinfo{number}{1} (\bibinfo{year}{2019}), \bibinfo{pages}{1--8}.
\newblock


\bibitem[Taniguchi and Kaufman(2021)]%
        {taniguchi2021family}
\bibfield{author}{\bibinfo{person}{Hiromi Taniguchi} {and} \bibinfo{person}{Gayle Kaufman}.} \bibinfo{year}{2021}\natexlab{}.
\newblock \showarticletitle{Family, Collectivism, and Loneliness from a Cross-Country Perspective}.
\newblock \bibinfo{journal}{\emph{Applied Research in Quality of Life}} (\bibinfo{year}{2021}), \bibinfo{pages}{1--27}.
\newblock


\bibitem[Twenge et~al\mbox{.}(2019)]%
        {twenge2019less}
\bibfield{author}{\bibinfo{person}{Jean~M Twenge}, \bibinfo{person}{Brian~H Spitzberg}, {and} \bibinfo{person}{W~Keith Campbell}.} \bibinfo{year}{2019}\natexlab{}.
\newblock \showarticletitle{Less in-person social interaction with peers among US adolescents in the 21st century and links to loneliness}.
\newblock \bibinfo{journal}{\emph{Journal of Social and Personal Relationships}} \bibinfo{volume}{36}, \bibinfo{number}{6} (\bibinfo{year}{2019}), \bibinfo{pages}{1892--1913}.
\newblock


\bibitem[Vanhalst et~al\mbox{.}(2015)]%
        {vanhalst2015lonely}
\bibfield{author}{\bibinfo{person}{Janne Vanhalst}, \bibinfo{person}{Bart Soenens}, \bibinfo{person}{Koen Luyckx}, \bibinfo{person}{Stijn Van~Petegem}, \bibinfo{person}{Molly~S Weeks}, {and} \bibinfo{person}{Steven~R Asher}.} \bibinfo{year}{2015}\natexlab{}.
\newblock \showarticletitle{Why do the lonely stay lonely? Chronically lonely adolescents’ attributions and emotions in situations of social inclusion and exclusion.}
\newblock \bibinfo{journal}{\emph{Journal of Personality and Social Psychology}} \bibinfo{volume}{109}, \bibinfo{number}{5} (\bibinfo{year}{2015}), \bibinfo{pages}{932}.
\newblock


\bibitem[Vickerstaff et~al\mbox{.}(2019)]%
        {vickerstaff2019methods}
\bibfield{author}{\bibinfo{person}{Victoria Vickerstaff}, \bibinfo{person}{Rumana~Z Omar}, {and} \bibinfo{person}{Gareth Ambler}.} \bibinfo{year}{2019}\natexlab{}.
\newblock \showarticletitle{Methods to adjust for multiple comparisons in the analysis and sample size calculation of randomised controlled trials with multiple primary outcomes}.
\newblock \bibinfo{journal}{\emph{BMC medical research methodology}} \bibinfo{volume}{19}, \bibinfo{number}{1} (\bibinfo{year}{2019}), \bibinfo{pages}{1--13}.
\newblock


\bibitem[Wagner et~al\mbox{.}(2015)]%
        {wagner2015s}
\bibfield{author}{\bibinfo{person}{Claudia Wagner}, \bibinfo{person}{David Garcia}, \bibinfo{person}{Mohsen Jadidi}, {and} \bibinfo{person}{Markus Strohmaier}.} \bibinfo{year}{2015}\natexlab{}.
\newblock \showarticletitle{It's a man's Wikipedia? Assessing gender inequality in an online encyclopedia}.
\newblock \bibinfo{journal}{\emph{Proceedings of the international AAAI conference on web and social media}} \bibinfo{volume}{9}, \bibinfo{number}{1}, \bibinfo{pages}{454--463}.
\newblock


\bibitem[Wahle et~al\mbox{.}(2016)]%
        {wahle2016mobile}
\bibfield{author}{\bibinfo{person}{Fabian Wahle}, \bibinfo{person}{Tobias Kowatsch}, \bibinfo{person}{Elgar Fleisch}, \bibinfo{person}{Michael Rufer}, \bibinfo{person}{Steffi Weidt}, {et~al\mbox{.}}} \bibinfo{year}{2016}\natexlab{}.
\newblock \showarticletitle{Mobile sensing and support for people with depression: a pilot trial in the wild}.
\newblock \bibinfo{journal}{\emph{JMIR mHealth and uHealth}} \bibinfo{volume}{4}, \bibinfo{number}{3} (\bibinfo{year}{2016}), \bibinfo{pages}{e5960}.
\newblock


\bibitem[Wang et~al\mbox{.}(2020)]%
        {wang2020social}
\bibfield{author}{\bibinfo{person}{Weichen Wang}, \bibinfo{person}{Shayan Mirjafari}, \bibinfo{person}{Gabriella Harari}, \bibinfo{person}{Dror Ben-Zeev}, \bibinfo{person}{Rachel Brian}, \bibinfo{person}{Tanzeem Choudhury}, \bibinfo{person}{Marta Hauser}, \bibinfo{person}{John Kane}, \bibinfo{person}{Kizito Masaba}, \bibinfo{person}{Subigya Nepal}, {et~al\mbox{.}}} \bibinfo{year}{2020}\natexlab{}.
\newblock \showarticletitle{Social sensing: assessing social functioning of patients living with schizophrenia using mobile phone sensing}.
\newblock \bibinfo{journal}{\emph{Proceedings of the 2020 CHI conference on human factors in computing systems}}, \bibinfo{pages}{1--15}.
\newblock


\bibitem[Wang et~al\mbox{.}(2023)]%
        {wang2023detecting}
\bibfield{author}{\bibinfo{person}{Zhiyuan Wang}, \bibinfo{person}{Maria~A Larrazabal}, \bibinfo{person}{Mark Rucker}, \bibinfo{person}{Emma~R Toner}, \bibinfo{person}{Katharine~E Daniel}, \bibinfo{person}{Shashwat Kumar}, \bibinfo{person}{Mehdi Boukhechba}, \bibinfo{person}{Bethany~A Teachman}, {and} \bibinfo{person}{Laura~E Barnes}.} \bibinfo{year}{2023}\natexlab{}.
\newblock \showarticletitle{Detecting Social Contexts from Mobile Sensing Indicators in Virtual Interactions with Socially Anxious Individuals}.
\newblock \bibinfo{journal}{\emph{Proceedings of the ACM on Interactive, Mobile, Wearable and Ubiquitous Technologies}} \bibinfo{volume}{7}, \bibinfo{number}{3} (\bibinfo{year}{2023}), \bibinfo{pages}{1--26}.
\newblock


\bibitem[Weeks and Asher(2012)]%
        {weeks2012loneliness}
\bibfield{author}{\bibinfo{person}{Molly~Stroud Weeks} {and} \bibinfo{person}{Steven~R Asher}.} \bibinfo{year}{2012}\natexlab{}.
\newblock \showarticletitle{Loneliness in childhood: Toward the next generation of assessment and research}.
\newblock \bibinfo{journal}{\emph{Advances in child development and behavior}}  \bibinfo{volume}{42} (\bibinfo{year}{2012}), \bibinfo{pages}{1--39}.
\newblock


\bibitem[Xu et~al\mbox{.}(2023)]%
        {Xu2023Globem}
\bibfield{author}{\bibinfo{person}{Xuhai Xu}, \bibinfo{person}{Xin Liu}, \bibinfo{person}{Han Zhang}, \bibinfo{person}{Weichen Wang}, \bibinfo{person}{Subigya Nepal}, \bibinfo{person}{Yasaman Sefidgar}, \bibinfo{person}{Woosuk Seo}, \bibinfo{person}{Kevin~S. Kuehn}, \bibinfo{person}{Jeremy~F. Huckins}, \bibinfo{person}{Margaret~E. Morris}, \bibinfo{person}{Paula~S. Nurius}, \bibinfo{person}{Eve~A. Riskin}, \bibinfo{person}{Shwetak Patel}, \bibinfo{person}{Tim Althoff}, \bibinfo{person}{Andrew Campbell}, \bibinfo{person}{Anind~K. Dey}, {and} \bibinfo{person}{Jennifer Mankoff}.} \bibinfo{year}{2023}\natexlab{}.
\newblock \showarticletitle{GLOBEM: Cross-Dataset Generalization of Longitudinal Human Behavior Modeling}.
\newblock \bibinfo{journal}{\emph{Proc. ACM Interact. Mob. Wearable Ubiquitous Technol.}} \bibinfo{volume}{6}, \bibinfo{number}{4}, Article \bibinfo{articleno}{190} (\bibinfo{date}{jan} \bibinfo{year}{2023}), \bibinfo{numpages}{34}~pages.
\newblock
\urldef\tempurl%
\url{https://doi.org/10.1145/3569485}
\showDOI{\tempurl}


\bibitem[Xu et~al\mbox{.}(2021)]%
        {xu2021understanding}
\bibfield{author}{\bibinfo{person}{Xuhai Xu}, \bibinfo{person}{Jennifer Mankoff}, {and} \bibinfo{person}{Anind~K Dey}.} \bibinfo{year}{2021}\natexlab{}.
\newblock \showarticletitle{Understanding practices and needs of researchers in human state modeling by passive mobile sensing}.
\newblock \bibinfo{journal}{\emph{CCF Transactions on Pervasive Computing and Interaction}} \bibinfo{volume}{3}, \bibinfo{number}{4} (\bibinfo{year}{2021}), \bibinfo{pages}{344--366}.
\newblock


\bibitem[Yan et~al\mbox{.}(2013)]%
        {yan2013smartphone}
\bibfield{author}{\bibinfo{person}{Zhixian Yan}, \bibinfo{person}{Jun Yang}, {and} \bibinfo{person}{Emmanuel~Munguia Tapia}.} \bibinfo{year}{2013}\natexlab{}.
\newblock \showarticletitle{Smartphone bluetooth based social sensing}.
\newblock \bibinfo{journal}{\emph{Proceedings of the 2013 ACM conference on Pervasive and ubiquitous computing adjunct publication}}, \bibinfo{pages}{95--98}.
\newblock


\bibitem[Y{\"u}r{\"u}r et~al\mbox{.}(2014)]%
        {yurur2014context}
\bibfield{author}{\bibinfo{person}{{\"O}zg{\"u}r Y{\"u}r{\"u}r}, \bibinfo{person}{Chi~Harold Liu}, \bibinfo{person}{Zhengguo Sheng}, \bibinfo{person}{Victor~CM Leung}, \bibinfo{person}{Wilfrido Moreno}, {and} \bibinfo{person}{Kin~K Leung}.} \bibinfo{year}{2014}\natexlab{}.
\newblock \showarticletitle{Context-awareness for mobile sensing: A survey and future directions}.
\newblock \bibinfo{journal}{\emph{IEEE Communications Surveys \& Tutorials}} \bibinfo{volume}{18}, \bibinfo{number}{1} (\bibinfo{year}{2014}), \bibinfo{pages}{68--93}.
\newblock


\bibitem[Zhao et~al\mbox{.}(2018)]%
        {zhao2018impact}
\bibfield{author}{\bibinfo{person}{Chenyue Zhao}, \bibinfo{person}{Feng Wang}, \bibinfo{person}{Xudong Zhou}, \bibinfo{person}{Minmin Jiang}, {and} \bibinfo{person}{Therese Hesketh}.} \bibinfo{year}{2018}\natexlab{}.
\newblock \showarticletitle{Impact of parental migration on psychosocial well-being of children left behind: a qualitative study in rural China}.
\newblock \bibinfo{journal}{\emph{International journal for equity in health}} \bibinfo{volume}{17}, \bibinfo{number}{1} (\bibinfo{year}{2018}), \bibinfo{pages}{1--10}.
\newblock


\bibitem[Zhou et~al\mbox{.}(2018)]%
        {zhou2018missing}
\bibfield{author}{\bibinfo{person}{Yuchao Zhou}, \bibinfo{person}{Suparna De}, \bibinfo{person}{Wei Wang}, \bibinfo{person}{Ruili Wang}, {and} \bibinfo{person}{Klaus Moessner}.} \bibinfo{year}{2018}\natexlab{}.
\newblock \showarticletitle{Missing data estimation in mobile sensing environments}.
\newblock \bibinfo{journal}{\emph{IEEE Access}}  \bibinfo{volume}{6} (\bibinfo{year}{2018}), \bibinfo{pages}{69869--69882}.
\newblock


\bibitem[Zou and Schiebinger(2018)]%
        {zou2018ai}
\bibfield{author}{\bibinfo{person}{James Zou} {and} \bibinfo{person}{Londa Schiebinger}.} \bibinfo{year}{2018}\natexlab{}.
\newblock \bibinfo{title}{AI can be sexist and racist—it’s time to make it fair}.
\newblock
\newblock


\end{thebibliography}

\newpage 
\appendix

\begin{small}

\begin{table*}[h]
  \caption{Summary tables of sensors and extracted sensor features by authors of the paper that conducted feature processing. The sensor features are computed by aggregating ten minutes of sensor data.}
  \label{tab:sensor_features}
  \begin{tabular}{l p{2.1cm} p{10cm}}

    \cellcolor[HTML]{EDEDED}\textbf{Sensor} &
    \cellcolor[HTML]{EDEDED}\textbf{\# of Features} &
    \cellcolor[HTML]{EDEDED} \textbf{Features and Description}
    \\

    Location (GPS) & 3 & Radius of gyration, the sum of distance covered, altitude \\
    \arrayrulecolor{Gray}
    \midrule
     
Bluetooth (low energy) & 5 & Number of devices connected, mean rssi (received signal strength indicator), max rssi, min rssi, std rssi   \\
\arrayrulecolor{Gray}
    \midrule

   Bluetooth (Normal) & 5 & Number of devices connected, mean rssi (received signal strength indicator), max rssi, min rssi, std rssi   \\
   \arrayrulecolor{Gray}
    \midrule
   
    Wifi & 6 & Number of devices connected to the device wifi hotspot, if the device is connected to a wifi,  wifi mean rssi (received signal strength indicator), max rssi, min rssi, std rssi \\
    \arrayrulecolor{Gray}
    \midrule
    
    Cellular gsm (2G) & 4 & Strength of the mobile signal as defined by mean, max, min, std \\
    
    \arrayrulecolor{Gray}
    \midrule

    Cellular wcdma (3G) & 4 & Strength of the mobile signal as defined by mean, max, min, std \\
    \arrayrulecolor{Gray}
    \midrule
    
    Cellular LTE (4G) & 4 & Strength of the mobile signal as defined by mean, max, min, std \\ 
    \arrayrulecolor{Gray}
    \midrule

    Notifications & 4 & notifications posted, notifications removed, notifications posted without duplicates, notifications removed without duplicates \\
    \arrayrulecolor{Gray}
    \midrule
    
    Proximity & 4 & Measures of the proximity sensor as mean, max, min, std \\
    \arrayrulecolor{Gray}
    \midrule
    
    Activity & 8 & Activities as classified from the accelerometer data by the Google Activity Recognition API \cite{googleActivityApi}; activity still, activity tilting, activity invehicle, activity onbicycle, activity onfoot, activity walking, activity running, activity unknown \\ 
    \arrayrulecolor{Gray}
    \midrule

    Steps & 2 &  step counted, steps detected \\
    \arrayrulecolor{Gray}
    \midrule
    
    Touch & 1 & Number of touch events \\
    \arrayrulecolor{Gray}
    \midrule
    
    Screen & 7 & User presence time, screen number of episodes, screen time total, screen time per episode,  screen time max episode, screen time min episode, screen time std episode \\ 
    \arrayrulecolor{Gray}
    \midrule
    
    Apps Usage & 48 & Time spent using app categories. App categories as classified by the google playstore \cite{santani2018drinksense}. E.g.: Action apps, dating apps, music apps, puzzle apps, etc. \\
    \arrayrulecolor{Gray}
    \midrule
    
    
    Time & 2 & Hour of the day, day of the week \\
    \arrayrulecolor{Gray2}
    \midrule
    
  \end{tabular}
\end{table*}

\end{small}

\end{document}